\documentstyle[preprint,prb,aps,epsf,eqsecnum]{revtex}
\begin{document}
\tightenlines
\title{Single Particle and Fermi Liquid Properties of
$^3$He\,--$^4$He Mixtures:\\
A Microscopic Analysis}
\author{E.~Krotscheck$^{1}$, J. Paaso$^2$,
M.~Saarela$^2$, K. Sch\"orkhuber$^1$, and R. Zillich$^1$}
\address{$^1$Institut f\"ur Theoretische Physik,
Johannes Kepler Universit\"at, A 4040 Linz, Austria}
\address{$^2$Department of Physical Sciences, Theoretical
Physics,
University of Oulu, FIN-90570 Oulu, Finland}
\maketitle

\begin{abstract}

We calculate microscopically the properties of the dilute
$^3$He component in a $^3$He\,--$^4$He-mixture. These depend on both,
the dominant interaction between the impurity atom and the background,
and the Fermi liquid contribution due to the interaction between the
constituents of the $^3$He component.
We first calculate the dynamic structure function of a $^3$He impurity
atom moving in $^4$He. From that we obtain the excitation spectrum and
the momentum dependent effective mass. The pole strength of this excitation
mode is strongly reduced from the free particle value in agreement
with experiments; part of the strength is distributed over high
frequency excitations. Above $k\geq 1.7$\AA$^{-1}$ the motion of
the impurity is damped due to the decay into a roton and a low energy
impurity mode.
Next we determine the Fermi--Liquid interaction between $^3$He atoms and
calculate the pressure-- and concentration dependence of the effective
mass, magnetic susceptibility, and the $^3$He\,--$^3$He scattering phase
shifts. The calculations are based on a dynamic theory that uses, as
input, effective interactions provided by the Fermi hypernetted--chain
theory. The relationship between both theories is discussed.
Our theoretical effective masses agree well with recent
measurements by Yorozu {\it et al.\/} (Phys. Rev. B {\bf 48}, 9660
(1993)) as well as those by R. Simons and R. M. Mueller (Czekoslowak
Journal of Physics Suppl. {\bf 46}, 201 (1996)), but our analysis
suggests a new extrapolation to the zero-concentration limit.
With that effective mass we also find a good
agreement with the measured Landau parameter $F_0^a$.


\end{abstract}

\section{INTRODUCTION}

Many ground state properties of $^3$He\,--$^4$He mixtures, in particular
the energetics of the system and the local structure, are today quite
well understood both experimentally\cite{OuY} and theoretically from a
microscopic point of view.\cite{MixMonster} With ``microscopic'' we
mean that one postulates no more knowledge than the empirical
Hamiltonian
\begin{equation}
H = -\sum_\alpha\sum_{i=1}^{N_\alpha}
{\hbar^2\over 2m_\alpha}\nabla_i^2 +{1\over 2}\sum_{\alpha\beta}
\mathop{{\sum}'}_{i,j}^{N_\alpha, N_\beta}
V(|{\bf r}_i^{(\alpha)}- {\bf r}_j^{(\beta)}|)
\label{hamiltonian}
\end{equation}
that contains only a local two-body interaction (recent work uses most
frequently the Aziz\cite{Aziz} interaction) and the masses of the two
species of particles.  This paper builds on our previous studies of
ground state properties of $^3$He\,--$^4$He mixtures\cite{MixMonster}
and properties of single $^3$He impurities\cite{SKJLTP} in $^4$He.
These calculations have produced an overall accuracy of the energetics
of the order of a few tenths of a percent for all experimentally
accessible densities and concentrations.

To specify our notation we use in the following Greek subscripts
$\alpha, \beta,\ldots \in \{3,4\}$ to refer to the {\it particle
species\/} (a $^3{\rm He}$ or a $^4{\rm He}$ particle), and Latin
subscripts $i, j,\ldots$ as in the ${\bf r}_i$ to refer to the
individual particles. The prime on the summation symbol in
Eq.~(\ref{hamiltonian}) indicates that no two pairs $(i,\alpha)$,
$(j,\beta)$ can be the same.  The number of particles of each species
is $N_\alpha$, and $N = N_3 + N_4$ is the total number of particles in
the system. In terms of the $^3$He  concentration $x$ we
have
\begin{equation}
        N_3 = x N,\qquad N_4=(1-x)N
\end{equation}
and the corresponding densities $\rho^{(\alpha)} = N_\alpha/\Omega$ are
inversely proportional to the volume $\Omega$ occupied by the whole fluid.

The overall most successful microscopic theory of the helium liquids
is the Jastrow-Feenberg variational method
\cite{FeenbergBook,Chuckreview} which uses a variational {\it
ansatz\/} for the ground state wave function of the form
\begin{eqnarray}
\Psi_0(\{{\bf r}_i^{(\alpha)}\})
        &&=e^{{1\over 2} U(\{{\bf r}_i^{(\alpha)}\})}
        \Phi_0(\{{\bf r}_i^{(3)}\})\nonumber\\
        U(\{{\bf r}_i^{(\alpha)}\})&& = {1\over 2!}\sum_{\alpha\beta}
        \mathop{{\sum}'}_{i,j}^{N_\alpha, N_\beta}
        u^{(\alpha\beta)}({\bf r}_i,{\bf r}_j)
        +{1\over 3!}\sum_{\alpha\beta\gamma}
        \mathop{{\sum}'}_{i,j,k}^{N_\alpha,N_\beta,N_\gamma}
        u^{(\alpha\beta\gamma)}
        ({\bf r}_i,{\bf r}_j,{\bf r}_k)\ .
\label{groundstate}
\end{eqnarray}
Here $\Phi_0(\{{\bf r}_i^{(3)}\})$ is the Slater determinant of plane
waves ensuring the antisymmetry of the Fermion component of the wave
function. The functions $u^{(\alpha\beta)}({\bf r}_i,{\bf r}_j)$ and
$u^{(\alpha\beta\gamma)}({\bf r}_i,{\bf r}_j,{\bf r}_k)$ are the pair
and triplet correlations; the species superscripts determine the type
of correlation. An essential part of the method is the optimization of
the ground state correlations by the variational principles
\cite{PPA1,PPA2,EKthree}
\begin{equation}
        {\delta E_0\over
                \delta u^{(\alpha\beta)}({\bf r}_i,{\bf r}_j)}=0\,,
\qquad
        {\delta E_0\over \delta u^{(\alpha\beta\gamma)}
        ({\bf r}_i,{\bf r}_j,{\bf r}_k)} = 0\; ,
\label{eulers}
\end{equation}
where
\begin{equation}
        E_0 = {\langle\Psi_0 | H | \Psi_0\rangle
        \over\langle\Psi_0 | \Psi_0\rangle}
\end{equation}
is the variational energy expectation value. Details of the procedure,
and the necessary working formulas have been discussed at length in
Ref. \onlinecite{MixMonster}.

\section{DYNAMICS OF A SINGLE $^3$HE ATOM}
\label{single}

This section briefly reviews our method \cite{MikkoIsrael,SKJLTP} to
calculate the dynamics of an impurity atom in the host liquid $^4$He
in its ground state with the wave function $\Psi({\bf r}_1,...,{\bf
r}_N)$.  The Jastrow-Feenberg wave function for the ground state of a
single impurity atom in $^4$He is easily derived from the wave
function (\ref{groundstate}) by setting $N_3 = 1$ and the Slater
determinant $\Phi_0({\bf r}_0)$ equal to unity,
\begin{eqnarray}
        \Psi^{(3)} ({\bf r}_0,{\bf r}_1,...,{\bf r}_N)
        = \exp {1\over2} \Bigl[ \sum_{j=1}^N u^{(34)}({\bf r}_0,{\bf r}_j) 
        +{1\over2!}\mathop{\sum_{j,k=1}}_{j\ne k}^N
        u^{(344)}({\bf r}_0,{\bf r}_j,{\bf r}_k)
        \Bigr]\Psi({\bf r}_1,...,{\bf r}_N)\ .
\label{waveI}
\end{eqnarray}
The coordinate ${\bf r}_0$ refers to the $^3$He atom and ${\bf
r}_1\ldots {\bf r}_N$ to the $^4$He atoms.  Results for ground state
calculations with this wave function have been reported
elsewhere.\cite{SKJLTP} They are consistent with the low concentration
limit of the calculations of Ref. \onlinecite{MixMonster} and in
quantitative agreement with experiments.

The dynamics of an impurity atom is determined by its response to a
weak, external time--dependent perturbation $U_{\rm ext}({\bf r}_0;t)$.
A natural generalization of the wave function (\ref{waveI}) for a
moving impurity atom is to allow for {\it time-dependent\/}
correlations. The {\it kinematic\/} and {\it dynamic\/} correlations
are separated by writing the wave function in the form
\begin{equation}
        \Phi (t) = e^{-iE_{N+1} t / \hbar }\;
        \Psi^{(3)} ({\bf r}_0,{\bf r}_1,...{\bf r}_N;t)
        /\sqrt{\left\langle \Psi^{(3)}(t)\mid\Psi^{(3)}(t)\right\rangle}\ ,
\end{equation}
where $E_{N+1}$ is the variational ground state energy of the $N+1$
particle system, and $\Psi^{(3)} ({\bf r}_0,{\bf r}_1,...{\bf r}_N;t)$
contains the time--dependent correlations,
\begin{equation}
        \Psi^{(3)} ({\bf r}_0, {\bf r}_1,... {\bf r}_N;t)
        = \exp {1 \over 2}
        \Bigl[\delta u^{(3)}({\bf r}_0;t) +  \sum_{i=1}^N
        \delta u^{(34)}({\bf r}_0,{\bf r}_i;t) \Bigr]
        \Psi^{(3)}({\bf r}_0,{\bf r}_1,...,{\bf r}_N)\ .
\label{timedep}
\end{equation}
The time--dependent components of the wave function are determined
by an action principle, searching for a stationary value of the
action integral
\begin{equation}
        {\cal L} = \int\! dt \;
        \langle \Phi (t) \vert H^{(3)}+U_{\rm ext}({\bf r}_0;t)
        -i\hbar {\partial \over
        \partial t} \vert \Phi (t) \rangle\ ,
\label{L}
\end{equation}
where $H^{(3)}$ is the Hamiltonian of the impurity-background system,
obtained from Eq. (\ref{hamiltonian}) for $N_3 = 1$. 

We make two assumptions in the evaluation of the action
integral. First we require that the pair-- and triplet correlation
functions  in
the ground state are optimized. This is important because it
eliminates all contributions to the action integral (\ref{L}) that are
linear in the time-dependent correlation functions. Then we assume
that the perturbation is weak which allows us to keep only the
quadratic terms in $\delta u^{(3)}({\bf r}_0;t)$ and $\delta
u^{(34)}({\bf r}_0,{\bf r}_i;t)$. The expression (\ref{L}) simplifies
because the potential energy term commutes with the time--dependent
part of the wave function and we are left with second-order terms
originating from the kinetic energy and the time
derivative only.\cite{MikkoIsrael,SKJLTP}

The variation of the action integral with respect to $\delta
u^{(3)}({\bf r}_0;t)$ and $\delta u^{(34)}({\bf r}_0,{\bf r}_i;t)$ leads
to one- and two-particle continuity equations, respectively,
\cite{MikkoIsrael,SKJLTP}
\begin{eqnarray}
        & &\nabla_0 \cdot {\bf j}^{(3)} ( {\bf r}_0;t)
        + {\partial\over\partial t}\delta\rho^{(3)}({\bf r}_0;t)
        = {2\rho^{(3)}\over\hbar} U_{\rm ext}( {\bf r}_0;t)
\label{continuityI}\\
        & &\nabla_0 \cdot {\bf j}^{(34)} ({\bf r}_0, {\bf r}_1;t) +
        \nabla_1 \cdot {\bf J}^{(34)}({\bf r}_0,{\bf r}_1;t)
        +{\partial\over\partial t}
        \delta\rho^{(34)}({\bf r}_0,{\bf r}_1;t) 
        = {2\over\hbar}U_{\rm ext}({\bf
        r}_0;t) \rho^{(34)}({\bf r}_0,{\bf r}_1)\ .
\label{continuityII}
\end{eqnarray}
We have kept only the time dependent parts of the full
densities. The {\it transition currents\/} are defined in terms of the
fluctuating one-particle density and pair correlation function:
\begin{eqnarray}
        {\bf j}^{(3)}({\bf r}_0;t)
        &&= {\hbar \over 2m_3i}\biggl[ \nabla_0\delta\rho^{(3)}({\bf r}_0;t)
        -  \int \! d^3r_1 \; \delta u^{(34)}({\bf r}_0,{\bf r}_1;t)
                \nabla_0 \rho^{(34)}({\bf r}_0,{\bf r}_1) 
        \biggr],
\label{jone}\\
        {\bf j}^{(34)}({\bf r}_0, {\bf r}_1;t)
        &&={1\over\rho^{(3)}}{\bf j}^{(3)}({\bf r}_0;t)
        \rho^{(34)}({\bf r}_0,{\bf r}_1)
        +{\hbar \over 2m_3i}
        \Biggl[ \rho^{(34)}({\bf r}_0,{\bf r}_1)
                \nabla_0 \delta u^{(34)}({\bf r}_0,{\bf r}_1;t) 
\nonumber\\
        +\int \! d^3r_2&&
        \Bigl[ \rho^{(344)}({\bf r}_0,{\bf r}_1,{\bf r}_2) 
        - {1\over\rho^{(3)}}\rho^{(34)}({\bf r}_0,{\bf r}_1)
        \rho^{(34)}({\bf r}_0,{\bf r}_2)\Bigr]
        \nabla_2 \delta u^{(34)}({\bf r}_0,{\bf r}_2;t)
        \Biggr]\ ,
\label{jtwo}\\
        {\bf J}^{(34)} ({\bf r}_0, {\bf r}_1;t)
        &&= {\hbar \over 2m_4i}\rho^{(34)}({\bf r}_0,{\bf r}_1)
        \nabla_1 \delta u^{(34)}({\bf r}_0,{\bf r}_1;t)\ .
\label{curr}
\end{eqnarray}
>From the calculation of the ground state properties of the impurity we
need here the impurity-background pair and triplet distribution functions 
$\rho^{(34)}({\bf r}_0,{\bf r}_1)$ and 
$\rho^{(344)}({\bf r}_0,{\bf r}_1,{\bf r}_2)$, respectively.

We assume that an infinitesimal external potential drives the system
with a given frequency and wavelength, and the system responds with
a density fluctuations of the same frequency and wavelength.
This determines the linear response function
\begin{equation}
        \chi^{(3)}(k,\omega)={\delta\rho^{(3)}(k,\omega)
        \over \rho^{(3)}\tilde U_{\rm ext}(k,\omega)}\ .
\end{equation}
The inverse of the response function can be calculated from the
Fourier transform of the continuity equations.  The Fourier transform
of the density fluctuations is defined as follows
\begin{eqnarray}
        \delta\rho^{(3)}(k,\omega)&=&\int \! d^3r_0 \, dt\; 
         e^{-i({\bf k}\cdot{\bf r}_0-\omega t)}
        \delta\rho^{(3)}({\bf r}_0;t)\ ,
\label{deltarho}
\end{eqnarray}
and analogously for the external potential.

The poles of
$\chi^{(3)}(k,\omega)$ determine the elementary excitations; their
dispersion relation is obtained from the first continuity equation by
setting $\tilde U_{\rm ext}(k,\omega)=0$. This leads to the implicit
equation
\begin{equation}
        \hbar \omega = {\hbar^2k^2\over 2m_3} 
        + \Sigma^{(3)} (k,\omega) 
\label{conI}
\end{equation}
with the self-energy
\begin{equation}
        \Sigma^{(3)} (k,\omega) =  {\hbar^2 \over 2m_3} 
        \int {d^3 p \over (2
        \pi)^3 \rho^{(4)}} \; {{\bf k}\cdot{\bf p} \; S^{(34)}(p)
        \beta^{(34)}_{{\bf k},\omega}({\bf p})\over
        \Bigl[\hbar\omega-t^{(3)}({\bf k}
        +{\bf p})-\epsilon^{(4)}(p)\Bigr]}\ .
\label{self}
\end{equation}
Here $S^{(34)}(p)$ is the $^3$He\,--$^4$He structure function,
$t^{(3)}(k)$ the kinetic energy of the impurity and
$\epsilon^{(4)}(p)$ the background phonon-roton spectrum. The function
$\beta^{(34)}_{{\bf k},\omega}({\bf p})$ will be defined below.

The linear response function has then a simple form
\begin{equation}
        \chi^{(3)}(k,\omega)=
         {1\over \hbar\omega-t^{(3)}(k)-\Sigma^{(3)} (k,\omega)+i\eta}
        -{1\over \hbar\omega+t^{(3)}(k)+\Sigma^{(3)} (k,\omega)+i\eta}
        ,~~\eta\rightarrow 0+
         \; .
\label{linres}
\end{equation}
This is the density--density response function of the $^3$He component
in the dilute limit. 
The response function is related to the Green's function
through\cite{FetterWalecka}
\begin{equation}
        \chi^{(3)}(k,\omega) = -i\int {d^3 q \; d(\hbar\omega^\prime)
        \over (2\pi)^4}\;  G(q,\omega^\prime)\; G({\bf k}+{\bf
        q},\omega^\prime +\omega) \; ,
\label{greenfree}
\end{equation}
where
\begin{equation}
        G(q,\omega) = {1-n_{\bf q}\over \hbar\omega-t^{(3)}(q)
        -\Sigma^{(3)}(q,\omega)+i\eta}
        + {n_{\bf q}\over \hbar\omega-t^{(3)}(q)
        -\Sigma^{(3)}(q,\omega)-i\eta} 
        ,~~\eta\rightarrow 0+
\end{equation}
and $n_{\bf q}$ is the Fermi distribution. In the dilute limit, we
have $n_{\bf q} = \delta_{{\bf q},0}$ and Eq. (\ref{greenfree})
reduces to Eq. (\ref{linres}) or, vice versa, the first part of
Eq. (\ref{linres}) can also be regarded as the single--particle
Green's function.

The imaginary part of the linear
response function defines the dynamic structure function
\begin{equation}
        S(k,\omega)= -{1\over\pi} \Im m\,[\chi^{(3)}(k,\omega)] \; ,
\label{dynstru}
\end{equation}
which can be 
measured in scattering experiments.

The contributions of the elementary excitations can be separated out
as a delta-function,
\begin{equation}
        S(k,\omega)=
        Z(k)\delta(\hbar\omega-\hbar\omega_p)+S_{\rm m}(k,\omega)\ . 
\label{pole}
\end{equation}
Here $\hbar\omega_p$ is the solution of Eq. (\ref{conI}) and
$S_{\rm m}(k,\omega)$ is the contribution of the modes that make the
self energy complex. This becomes possible when the denominator in
Eq. (\ref{self}) becomes positive for some value of ${\bf p}$,
\begin{equation}
        \max_{\bf p}\left[
        \hbar\omega-t^{(3)}({\bf k}+{\bf p})-\epsilon^{(4)}(p)
        \right]~>~0\ .
\label{denom}
\end{equation}
If the condition (\ref{denom}) is satisfied, then it is kinematically
possible that the $^3$He impurity loses energy by emitting a
phonon-roton mode $\epsilon^{(4)}(p)$ while making a transition into a
low-energy impurity mode $t^{(3)}({\bf k}+{\bf p})$.  The strength of
the pole $Z(k)$ can be evaluated from the derivative of the self
energy,
\begin{equation}
        Z(k)=\left[1-\left.{d \Sigma^{(3)} (k,\omega)\over
        d(\hbar\omega)}\right|_{\omega=\omega_p }\right]^{-1}\ .
\label{strength}
\end{equation}

The second continuity equation (\ref{continuityII}) can be written in
the Fourier space and formulated as a linear integral equation for
$\beta^{(34)}_{{\bf k},\omega}({\bf p})$, which is related to
the fluctuating pair correlation function as discussed in the
Appendix. The integral equation sums ladder diagrams up
to infinite order,
\begin{equation}
        \beta^{(34)}_{{\bf k},\omega}({\bf p})
        =\hbar\omega \; \frac{{\bf k}\cdot{\bf p}}{k^2}\; 
        \frac{S^{(34)}(p)}{S^{(44)}(p)}
        -\int{d^3q\over (2\pi)^3 \rho^{(4)}} \; 
        \frac{\beta^{(34)}_{{\bf k},\omega}({\bf q})
        K_{{\bf k},\omega}({\bf p},{\bf q})}
        {\Bigl[\hbar\omega-t^{(3)}({\bf k}+{\bf q})
        -\epsilon^{(4)}(q)\Bigr]}
\label{conII}
\end{equation}
with the kernel
\begin{eqnarray}
        K_{{\bf k},\omega}({\bf p},{\bf q})
        &=& S^{(44)}(q)\Biggl\{\left[\left(S^{(34)}(|{\bf p}-{\bf q}|)+1\right)
        \tilde u^{(344)}({\bf p}-{\bf q},-{\bf p},{\bf q})
        +S^{(34)}(|{\bf p}-{\bf q}|)\right] 
\nonumber \\
        &\times&\left[\hbar\omega
        -\frac{({\bf k}+{\bf p})\cdot({\bf k}+{\bf q})}{p^2}
        t^{(3)}(p)\right]\Biggr\}
        -S^{(34)}(|{\bf p}-{\bf q}|)\frac{{\bf p}\cdot{\bf q}}{p^2}
        \epsilon^{(4)}(p)\ . 
\label{kernel}
\end{eqnarray}
Here we need the static structure function $S^{(44)}(k)$ of the
background and the triplet correlation function $\tilde u^{(344)}$
with one impurity. Details of the derivation of the above equation are
given in the Appendix.

The singularity structure of Eq. (\ref{conII}) is the same as that of
the self energy. For real frequencies $\omega$ the imaginary part of
$\beta^{(34)}_{{\bf k},\omega}({\bf q})$ is zero if the energy
denominator is negative for all values of ${\bf q}$.  Modes with an
energy $\hbar\omega$ high enough to satisfy the inequality
(\ref{denom}) can decay into a phonon-roton mode and the solution has a
non-zero imaginary part.

The long-wavelength limit of the excitation energy defines the
hydrodynamic effective mass $m^*_H$.
\begin{equation}
  \hbar \omega = {\hbar^2k^2\over 2m^*_H}\ , ~~{\rm when}~~ k\to 0\ .
\label{hydromass}
\end{equation}
Inserting this into Eq. (\ref{conI}) we get 
\begin{equation}
{m^*_H\over m_3} = {1\over 1-I}   
\label{mhydro}
\end{equation}
with
\begin{equation}
        I =  \lim_{k\rightarrow 0+}{1\over k^2}\int {d^3 p \over (2
        \pi)^3 \rho^{(4)}} \; {{\bf k}\cdot{\bf p} \;  S^{(34)}(p)\;
        \beta^{(34)}_{{\bf k},\omega_0}({\bf p})\over 
        t^{(3)}(p)+\epsilon^{(4)}(p)} \ ,
\label{massinte}
\end{equation}
where $\omega_0=\hbar k^2/2m^*_H$.
Using Eq. (\ref{hydromass}) we find that in the long-wavelength limit
the pole strength is inversely proportional to the effective mass,
\begin{equation}
        \lim_{k\rightarrow 0} Z(k) = {m_3\over m^*_H}\ . 
\label{zzero}
\end{equation}

In the so-called ``uniform limit approximation''\cite{FeenbergBook}
one neglects all {\it coordinate-space\/} products of two
functions, {\it i.e.\/} we approximate, for example,
$\rho^{(34)}({\bf r}_0,{\bf
r}_1)\delta u^{(34)}({\bf r}_0,{\bf r}_1) \approx
\rho^{(3)}\rho^{(4)}\delta u^{(34)}({\bf r}_0,{\bf r}_1)$, but {\it
convolution products\/} are retained. Then, $\beta^{(34)}_{{\bf
k},\omega_0}({\bf p})$ has a simple form\cite{SKJLTP}
\begin{equation}
        \beta^{(34)}_{{\bf k},\omega_0}({\bf p}) 
        = {\hbar^2 \over 2m_3}\;{\bf k}\cdot{\bf p}\;
        {S^{(34)}(p)\over S^{(44)}(p)}\ . 
\label{conIIa}
\end{equation}
This together with the equations (\ref{mhydro}) and (\ref{massinte})
gives the ``un-renormalized effective mass'' derived by
Owen.\cite{OwenBackflow}

\section{FERMI LIQUID INTERACTIONS}

Up to this point, we have considered only the interaction of one
single $^3$He atom with the host liquid. Further interesting effects
arise from the interaction between pairs of $^3$He atoms and the
specific dynamics imposed on the $^3$He component by the Pauli
principle.\cite{LaP} The most obvious manifestations of interactions
between the $^3$He atoms are magnetic properties\cite{BashkinMeye} and
corrections to the hydrodynamic mass.  Both
effects provide interesting problems from the point of view of
low-temperature experiments as well as microscopic many--body theory.

The dilute mixture is a particularly attractive system for the
theorist because many of the complicated exchange effects that obscure
the theory of pure $^3$He are negligible. A new effect is that the
interaction between $^3$He atoms is dominated by the exchange of
phonons through the host liquid\cite{BashkinMeye,BBP66,BBP} and
therefore depends not only on the relative momentum between the two
particles, but also on the momentum of each individual particle
relative to the $^4$He background.

Small perturbations of the ground state of a normal, interacting Fermi
liquid are described within Landau's Fermi Liquid
theory.\cite{LaP} The energy $E_0$ of the system is considered to be a
functional of the quasiparticle occupation numbers $n_{{\bf
k},\sigma}$, and the quasiparticle energies are, close to the ground
state, given by the variations of that energy with respect to $n_{{\bf
k},\sigma}$. Thus, the quasiparticle spectrum is
\begin{equation}
\epsilon^{(3)}({\bf k},\sigma)
 = {\delta E_0 \over \delta n_{{\bf k},\sigma}}\ .
\label{epsqp}
\end{equation}
For a symmetric ground state, the $\epsilon^{(3)}({\bf k},\sigma)$ are
spin--independent; we shall suppress the spin argument whenever
possible.

In the mixture, the single particle spectrum consists of two
parts. The first is the hydrodynamic interaction of a single impurity
atom with the background, leading to the ``hydrodynamic effective
mass'' $m^*_H$ as discussed in the preceding section. The second part
is due to the interaction with other $^3$He atoms. Hence, we can split
the quasiparticle spectrum (\ref{epsqp}) into two pieces,
\begin{equation}
\epsilon^{(3)}( k)
= \epsilon^{(3)}_H( k) +  \epsilon^{(3)}_{\rm QP}(k)\ ,
\end{equation}
Correspondingly, the total effective mass of the $^3$He particles is
\begin{equation}
{\hbar^2 k_F\over m^*} \equiv {\hbar^2 k_F\over m^*_H}
+ \left.{d\over dk}\epsilon^{(3)}_{\rm QP}( k)\right|_{k=k_F}\ ,
\label{QPmass}
\end{equation}
where $k_F$ is the Fermi momentum of the $^3$He component.

Interactions between the $^3$He atoms are, to the extent that the
$^3$He atoms remain close to the Fermi surface, described by Landau's
Fermi liquid (``quasiparticle'') interaction which is the second
variation
\begin{equation}
f_{ {\bf k} \sigma , {\bf k}' \sigma' }^{\rm var} =
{\delta \epsilon^{(3)}({\bf k},\sigma)\over \delta n_{{\bf k}', \sigma'}}
= \left.{\delta^2 E_0 \over
\delta n_{{\bf k}, \sigma} \delta n_{{\bf k}',
\sigma'}}\right|_{n_k^{(0)}}\ ,
\label{smallf}
\end{equation}
evaluated at the ground state momentum distribution $n_k^{(0)} =
\theta(k_F-k)$.
The quasiparticle interaction normally contains a spin-independent and
a spin-dependent part,
\begin{eqnarray}
f_{ {\bf k} \sigma , {\bf k}' \sigma' }^{\rm var}
&=&f_{{\bf k},{\bf k}'}^{\rm s} + f_{{\bf k},{\bf k}'}^{\rm a} \;
        \mathbf{\sigma} \cdot \mathbf{\sigma}' \, ,
\label{nonlocalqp}
\end{eqnarray}
which is also frequently written as
\begin{equation}
f_{{\bf k},{\bf k}'}^{\rm s (a)}  = {1 \over 2}
\left(f_{\uparrow ,\uparrow }^{\rm var} \pm
f_{\uparrow ,\downarrow }^{\rm var}\right) \, ;
\end{equation}
the $\mathbf{\sigma}$'s are Pauli spin matrices. Since the
quasiparticle interaction is to be taken for
$\left|{\bf k}\right|=\left|{\bf k}'\right| = k_F$, it
depends only on the angle between ${\bf k}$ and ${\bf k}'$,
and it is convenient to expand that dependence in Legendre
polynomials
\begin{equation}
f_ {{\bf k}, {\bf k}'}^{\rm s (a)}
= \sum_\ell f_\ell^{s (a)} P_\ell(\hat{\bf k} \cdot \hat{\bf k}')\, .
\label{expandf}
\end{equation}
The strength of the interaction relative to the kinetic energy is
measured by the dimensionless quantities
\begin{equation}
  F_\ell^{\rm s (a)} = N(0) \;  f_\ell^{\rm s (a)} = {\Omega \; m^* \; k_F
  \over \pi^2 \; \hbar^2 } \;  f_\ell^{\rm s (a)} \; ,
\label{Bigf}
\end{equation}
where $N(0)$ is the density of states at the Fermi surface.

At this point, the quasiparticle contribution to the effective mass is
normally related to the Landau parameter $f_1^s$ or $F_1^s$. To derive
this relationship, one must assume Galilean--invariance of the whole
liquid. As already observed by Baym and Pethick,\cite{BaymPethick}
the argument must be treated with caution here because, while
the whole mixture is Galilean invariant, the properties of the $^3$He
component are not invariant against motion relative the $^4$He
background. This becomes relevant when one attempts to treat the
phonon--mediated processes as an effective interaction between the
$^3$He atoms: As long as this interaction is assumed to be static, it
can depend only on the properties of the participating $^3$He
particles. However, in reality the interaction is dominated by the
dynamic exchange of phonons through the background liquid and the
motion of the interacting particles relative to that background
becomes important. The correct definition of the effective mass is
therefore the one of Eq. (\ref{QPmass}); we will explicitly see
the effect of the retardation of the quasiparticle interaction
in our numerical applications.

The magnetic response is determined by the change of the quasiparticle
energy when a magnetic field $H$ is applied.  Since the $^4$He
background does not couple to the spin degrees of freedom, the $\delta
\epsilon^{(3)}({\bf k},\sigma)$ depends, to first order in the magnetic field,
only on the spin---populations $n_{{\bf k},\sigma}$.  Hence the usual
relationship of Landau theory,
\begin{eqnarray}
        {\chi^*_{\sigma,\rm ideal} \over \chi_\sigma} 
        &= & {m_H^*\over m^*} \; 
        \left( 1 + { m^* \; k_F \over \pi^2 \;
        \hbar^2} \,\Omega\, f_0^{\rm a} \right)
        = {m_H^*\over m^*} \; \left( 1 + F_0^{\rm a}\right) \; ,
\label{SuszLandau}
\end{eqnarray}
between the $F_0^{\rm a}$ and the spin--susceptibility is
applicable. In order to separate the hydrodynamic ``backflow''
effect from the Fermi liquid interaction, we have defined here
$\chi^*_{\sigma,\rm ideal}$ as the Pauli susceptibility for
$^3$He particles with mass $m_H^*$.

\subsection{Variational Fermi Liquid Theory}

Landau's Fermi liquid theory is phenomenological in the sense that it
relates Fermi liquid parameters to physical observables, and derives
relationships between different physical observables, but it makes no
statements on how to compute these quantities.  Methods to calculate
the quasiparticle excitations and interactions from the underlying
Hamiltonian (\ref{hamiltonian}) of the system are provided by
microscopic many-body theory.

At the level of the variational theory, the single particle excitation
spectrum is calculated by allowing for an occupation of single
particle orbitals $n_{{\bf k},\sigma}$ in the Slater function $\Phi_0$
that is different from the Fermion ground state $n_k^{(0)}$.  The
graphical analysis of the functional derivatives has been carried out
in Ref. \onlinecite{CBF2}; the variational single-particle field
$u(k)$ has the general form

\begin{equation}
\epsilon^{(3)}(k,\sigma) = t^{(3)} (k) + u(k) + U_0 \, ,
\label{epsHF}
\end{equation}
where $U_0$ is a constant determined by the chemical potential, and
$u(k)$ a momentum dependent average field. In the present dilute case,
the average field can be written in the form of a Hartree-Fock field
\begin{equation}
u(k) = - ~{\rho^{(3)}\over 2}\int d^3 r \, W_{\rm eff}(r) \, j_0(r \,
          k) \,\ell(r \, k_F)
\label{uFHNC}
\end{equation}
where $\ell(x) = 3 ~ j_1(x)/x$ is the Fermi exchange function familiar
from Hartree-Fock theory, and $W_{\rm eff}(r)$ is a local effective
interaction which can be constructed diagrammatically by the summation
of (F)HNC diagrams in the mixture. \cite{MixMonster,CBF2}
For further reference, we also formulate the average field in
momentum space,
\begin{equation}
u(k) = - \int {d^3 q\over (2\pi)^3\rho^{(3)}}
\, n_{|{\bf q}-{\bf k}|,\sigma} \,
\tilde W_{\rm eff}(q)\,,
\label{pFHNC}
\end{equation}
where $\tilde W_{\rm eff}(q)$ is the dimensionless Fourier transform 
$\tilde W_{\rm eff}(q) = \rho^{(3)}\int d^3 r  W_{\rm eff}(r)
\exp(i \; {\bf k}\cdot{\bf r})$.

In the same approximation, the quasiparticle interaction can also be
expressed, up to an additive constant, in terms of the effective
interaction $W_{\rm eff}(r)$:
\begin{eqnarray}
f_{ {\bf k} \sigma , {\bf k}' \sigma' }^{\rm var}
&=&{1\over\Omega}\left[{\rm const.} -
\delta_{\sigma,\sigma'} \, 
       \tilde W_{\rm eff}(|{\bf k}-{\bf k'}|)\right]\nonumber\\
&=&f_{{\bf k},{\bf k}'}^{\rm s} + f_{{\bf k},{\bf k}'}^{\rm a} \;
        \mathbf{\sigma} \cdot \mathbf{\sigma}' \, .
\label{localqp}
\end{eqnarray}
The constant first term in Eq. (\ref{localqp}) comes from the
variation of $U_0$ with respect to the occupation number $n_{{\bf
k},\sigma}$; it is related to the compressibility of the $^3$He
component and is therefore not our concern now.

\subsection{Correlated Basis Functions}

It has been known for quite some time\cite{KSCP81} that the na\"\i ve
implementation of the FHNC/EL theory (or, for that matter, {\it any\/}
variational wave function of the type (\ref{groundstate})) to the
problem of pure $^3$He has a number of severe problems. One is that it
leads to an effective mass of {\it pure\/} $^3$He that is less than
one\cite{IndianMass} which is in sharp contrast to
experiments.\cite{Wheatley75,AHMS80} Another problem is that it
predicts an instability against spontaneous spin-polarization. The
physics of mixtures is, of course, entirely different from that of a
pure $^3$He liquid. But we will see that the technical approximations
implicit to the Jastrow-Feenberg wave function (\ref{groundstate})
lead to practically the same problems in the mixture.

The deficiencies of the variational theory are, as
we shall see more explicitly below, due to the fact that the
wave function (\ref{groundstate}) describes the {\it average\/} correlations
between particles quite well, whereas it is insensitive to the {\it
specifics\/} of the correlations in the vicinity of the Fermi surface.
The cure for the problem is known to be Correlated-Basis Functions
(CBF) theory.\cite{FeenbergBook} A complete, non-orthogonal basis of
the Hilbert space is built on the ground state wave function
(\ref{groundstate}) by defining
\begin{equation}
\Psi_m(\{{\bf r}_i^{(\alpha)}\})
        =e^{{1\over 2} U(\{{\bf r}_i^{(\alpha)}\})}
        \Phi_m(\{{\bf r}_i^{(3)}\})\ ,
\label{CBFbasis}
\end{equation}
where the $\left\{\Phi_m(\{{\bf r}_i^{(3)}\})\right\}$ is a complete
set of Slater determinants for the $^3$He component. Within the basis
$\left\{\Psi_m(\{{\bf r}_i^{(\alpha)}\})\right\}$, perturbative
expansions can be derived in much the same way as in conventional
Rayleigh-Schr\"odinger perturbation theory, and diagrammatic
structures like ring-, ladder-, or self--energy diagrams can be
defined and summed to infinite order. Note that perturbative
corrections are applied to the Fermion components only; this is
legitimate since the Jastrow-Feenberg wave function is in principle
exact for Bosons, whereas it is not exact for Fermions.

The general strategy of CBF theory has been described in detail in
many places;\cite{FeenbergBook} the basic results needed for our
analysis have been derived in Ref.  \onlinecite{rings}. Without going
into the details of that theory, we assert that there is a one-to-one
correspondence between the compound-diagrammatic elements of the FHNC
theory and that of ordinary, Green's functions based perturbation
theory. Loosely speaking, CBF theory can be interpreted as a
microscopic procedure for the calculation of static effective
interactions. For our purpose, the application of CBF theory can be
summarized simply by stating that, in the evaluation of the energy,
the {\it chain-diagrams\/} of FHNC theory are replaced by the {\it
ring-diagrams\/} of the RPA theory, using a diagrammatically defined
static effective interaction which we identify with the
``particle-hole interaction''.

Since we are predominantly interested in variations of the energy, we
can formulate the theory in terms of a Green's functions approach where
the effective interactions come from the FHNC analysis, examine the
relationships between the FHNC and the CBF approximation, and compare
the results of different theories.

In perturbative theories, single particle properties are
described by a complex {\it self-energy\/} $\Sigma(k,\omega)$; the
single particle spectrum $\varepsilon(k)$ is obtained from the
solution of the equation
\begin{equation}
\varepsilon(k) = t^{(3)}(k) + \Sigma(k,\varepsilon(k)/ \hbar)\ .
\label{selfen}
\end{equation}
If only single phonon coupling processes are considered, the self-energy
$\Sigma(k,E)$ is given in the so-called
G0W-approximation\cite{Hedin65,Rice65}
\begin{equation}
\Sigma(k,\omega) = i \int { d^3 q d(\hbar\omega^\prime)\over 
  (2\pi)^4\rho^{(3)}} \,
   G^{(0)}(\left|{\bf k}-{\bf q}\right|,\omega-\omega^\prime) \,
\tilde V_{\rm eff\/}(q,\omega^\prime)
\label{G0W}
\end{equation}
where
\begin{equation}
G^{(0)}(k,\omega) =
{1-n_{{\bf k},\sigma}\over \hbar\omega - t^{(3)}(k) + i\eta} 
+ {n_{{\bf k},\sigma}\over \hbar\omega - t^{(3)}(k) - i\eta} 
\label{green}
\end{equation}
is the free single-particle Green's function and
\begin{equation}
\tilde V_{\rm eff\/}(q,\omega) = 
\tilde V_{\rm p-h}^{(33)}(q) + \sum_{\alpha\beta}
\tilde V_{\rm p-h}^{(3\alpha)}(q) \, \chi^{(\alpha\beta)}(q,\omega) \,
\tilde V_{\rm p-h}^{(3\beta)}(q)
\label{VeffE}
\end{equation}
is the effective, {\it energy dependent\/} $^3$He\,--$^3$He interaction.
In Eq. (\ref{VeffE}), the $\tilde V_{\rm p-h}^{\alpha\beta}(q)$ are
the local, {\it particle-hole irreducible interactions,\/} and
$\chi^{(\alpha\beta)}(q,\omega)$ is the density-density response
matrix. In fact, the G0W approximation is simply the variation of an
RPA energy expectation value with respect to a single particle Green's
function.  The key quantities $\tilde V_{\rm p-h}^{\alpha\beta}(q)$
can in principle be defined diagrammatically in Green's functions
based theories, but it is impractical to calculate these interactions
without further approximations. CBF perturbation theory solves this
problem by giving unambiguous prescriptions\cite{MixMonster} for the
calculation of {\it static approximations\/} for these quantities.

To separate the ``hydrodynamic'' and the ``Fermionic'' component
of the self-energy, we rewrite the single-particle Green's function
as
\begin{eqnarray}
G^{(0)}(k,\omega) &=& {1\over \hbar\omega - t^{(3)}(k) + i\eta} 
+n_{{\bf k},\sigma} \left[-{1\over \hbar\omega - t^{(3)}(k) + i\eta} 
+ {1\over \hbar\omega - t^{(3)}(k) - i\eta}\right]\nonumber\\
&\equiv& G_H^{(0)}(k,\omega) + G_F^{(0)}(k,\omega)
\label{green2}
\end{eqnarray}
and the self-energy in the form
\begin{eqnarray}
\Sigma(k,\omega) &=& \Sigma_H(k,\omega) + \Sigma_F(k,\omega),\\
\Sigma_H(k,\omega) &=& i \int {d^3 q \, d(\hbar\omega^\prime)\over 
 (2\pi)^4\,\rho^{(3)}} \; G_H^{(0)}(\left|{\bf k}-{\bf q}\right|,
   \omega-\omega^\prime) \; \tilde V_{\rm eff\/}(q,\omega^\prime),
\label{SigmaH}\\
\Sigma_F(k,\omega) &=& i \int {d^3 q \, d(\hbar\omega^\prime)\over 
   (2\pi)^4 \,\rho^{(3)}} \; G_F^{(0)}(\left|{\bf k}-{\bf q}\right|,
   \omega-\omega^\prime)\; \tilde V_{\rm eff\/}(q,\omega^\prime)\ .
\label{SigmaF}
\end{eqnarray}
The ``hydrodynamic'' part (\ref{SigmaH}) of the self-energy leads, in
the dilute limit $\rho^{(3)}\rightarrow 0$, to an effective mass that
is identical\cite{SKJLTP} to the one obtained in the previous section
in the ``uniform limit approximation'' derived in Eqs. (\ref{mhydro})
and (\ref{conIIa}). This approximation is quantitatively not quite
sufficient, but it contains, at a semi-quantitative level, the basic
physics of impurity motion, in particular the coupling of the impurity
to the excitations of the host liquid. The method of time--dependent
correlation functions described in Sec. \ref{single} introduces a
systematic procedure to improve upon this approximation.

Let us now focus on the Fermion term, Eq. (\ref{SigmaF}).
The energy integration yields the compact form
\begin{equation}
\Sigma_F(k,\omega) = -\int {d^3 q\over (2\pi)^3\rho^{(3)}} \, 
n_{|{\bf q}-{\bf k}|,\sigma} \,
\tilde V_{\rm eff\/}(q,\omega-t^{(3)}({\bf k}-{\bf q})/ \hbar) \, .
\label{SigmaFinal}
\end{equation}

To calculate the spin-susceptibility we can directly apply
Eqs. (\ref{Bigf}) and (\ref{SuszLandau}) and arrive at
\begin{equation}
\Omega f_0^{\rm a} = -{1 \over 2\rho^{(3)}}
                \int\!\!\int {d\theta \; d\phi \over 4 \pi} 
                \; \sin(\theta) \; \tilde V_{\rm eff\/}(\vert 
                {\bf k}_{\rm F}-{\bf k'}_{\rm F} \vert ,0) \, .
\label{f0a}
\end{equation}

In principle, we would also need to calculate the variation of the
kinetic energy $t^{(3)}({\bf k}-{\bf q})$ with respect to the
quasiparticle occupation number $n_{{\bf k},\sigma}$, but in the
dilute limit of the mixture, this quantity is dominated by the
hydrodynamic backflow relative to the background and can, to good
approximation, be replaced by a free single--particle spectrum with an
effective mass $m_H^*$.

The comparison of Eqs. (\ref{f0a}) and (\ref{SigmaFinal}) demonstrates
the difference between the Landau effective mass that would be
calculated from the $\ell = 1$ equivalent to Eq. (\ref{f0a}), and the
momentum derivative $\left.d\varepsilon(k,\sigma)/dk\right|_{k=k_F}$.
In the latter case, one also includes the momentum dependence of the
energy appearing in the effective interaction $\tilde V_{\rm
eff\/}(q,\omega-t^{(3)}({\bf k}-{\bf q})/ \hbar)$.

We have now reached the point where the quasiparticle interaction is
expressed in terms of a local, but energy dependent effective
interaction very similar to the FHNC theory, {\it cf.\/}
Eq. (\ref{localqp}).  The relationship to Eq. (\ref{pFHNC}) is now
apparent: $\Sigma_F(k,\omega)$ could be rewritten as a Hartree-Fock
expression of the form (\ref{pFHNC}) if the effective interaction
$\tilde V_{\rm eff\/}(q,\omega)$ were energy independent. The question
that arises naturally is what the relationship between the {\it
energy-independent\/} effective interaction $\tilde W_{\rm eff}(q)$
and the {\it energy dependent\/} interaction $\tilde V_{\rm
eff}(q,\omega)$ is. We will address this question in the next
subsection.

\subsection{Connection between HNC and CBF Theories}

We here outline the general manipulations that lead from the
compound-diagrammatic elements of the (F)HNC theory to the Green's
functions expressions.  The construction of energy-independent, local
effective interactions is one of the key steps in the Feynman-diagram
based parquet-diagram theory\cite{parquet1,parquet2} that is needed to
show the equivalence to the HNC/EL theory, and we shall demonstrate
here that the relationship between (F)HNC and parquet theory reach
actually farther than expected.

For simplicity, we restrict our derivation to the case of two
interacting $^3$He impurities within the $^4$He host liquid.  In this
case, we need to retain only the 44-channel of the response
function. In this limit, the effective interaction is
\begin{eqnarray}
\tilde V_{\rm eff\/}(q,\omega) &=& 
\tilde V_{\rm p-h}^{(33)}(q) + 
\tilde V_{\rm p-h}^{(34)}(q) \; \chi^{(44)}(q,\omega) \;
\tilde V_{\rm p-h}^{(34)}(q)\nonumber\\
&=&\tilde V_{\rm p-h}^{(33)}(q) + 
\tilde V_{\rm p-h}^{(34)}(q) \;
 {\displaystyle 2 t^{(4)}(q)\over (\hbar\omega)^2 - (\epsilon^{(4)}(q))^2}
\; \tilde V_{\rm p-h}^{(34)}(q)\ .
\label{effints}
\end{eqnarray}
The prescription from parquet-theory to make this energy-dependent
interaction local is as follows: Construct the RPA static structure
function
\begin{eqnarray}
S^{(33)}_{\rm RPA}(q) &=& \int {d(\hbar\omega)\over 2\pi} \;
\Im m\left[\chi^{(33)}_0(q,\omega) + \chi^{(33)}_0(q,\omega)
\tilde V_{\rm eff\/}(q,\omega)\chi^{(33)}_0(q,\omega)\right]
\nonumber\\
&=& 1 -{1\over t^{(3)}(q)}\left[\tilde V_{\rm p-h}^{(33)}(q)
- 2 \; {\left[V_{\rm p-h}^{(33)}(q)\right]^2 S^{(44)}(q)\over
\left(t^{(3)}(q) + \epsilon^{(4)}(q)\right)^2}\right]\ , 
\end{eqnarray}
where
\begin{equation}
\chi^{(33)}_0(q,\omega) = {2 \; t^{(3)}(q)\over
(\hbar\omega)^2 - (t^{(3)}(q))^2}
\end{equation} 
is the response function of the non-interacting $^3$He gas. Also,
construct the {\it ladder approximation\/} for the same quantity in
terms of a different and yet unspecified local effective interaction,
say $\tilde V_{\rm L}(q)$
\begin{eqnarray}
S^{(33)}_{\rm ladder}(q) &=& \int {d(\hbar\omega)\over 2\pi} \; 
\Im m\left[\chi^{(33)}_0(q,\omega) + \chi^{(33)}_0(q,\omega)\;
\tilde V_{\rm L\/}(q)\chi^{(33)}_0(q,\omega)\right]\nonumber\\
&=& 1 - {1\over t^{(3)}(q)}\; \tilde V_{\rm L\/}(q)\ .
\end{eqnarray}
Now choose an average frequency $\bar\omega(q)$ such that these
two forms of the static structure function are
{\it identical\/} for
\begin{equation}
\tilde V_{\rm L\/}(q) =
\tilde V_{\rm eff\/}(q,\bar\omega)\ .
\end{equation}
One obtains
\begin{equation}
(\hbar\bar\omega)^2 =
-{\epsilon^{(4)}(q) \; (t^{(3)})^2(q)\over\epsilon^{(4)}(q)+2t^{(3)}(q)}\ .
\end{equation}
This is the generalization of the expression given in Ref.
\onlinecite{parquet2} to a dilute $^3$He\,--$^4$He mixture.
{\it Inserting $\hbar\bar\omega$ into the energy-dependent
effective interaction leads directly to the
FHNC approximation\/} (\ref{uFHNC}), in other words
\begin{equation}
\tilde W_{\rm eff}(q) = \tilde V_{\rm eff\/}(q,\bar\omega)
= \tilde V_{\rm L}(q)\ .
\label{equiv}
\end{equation}
This final part is of a somewhat technical nature and requires the
full (F)HNC formalism, in particular its use to calculate effective
interactions for use in CBF theory\cite{rings,MixMonster} and the
connections with the optimization problem; it is therefore skipped
here.

The equivalence (\ref{equiv}) is the second essential theoretical
result of this paper. The result is remarkable in the following sense:
The concept of an ``average energy'' has been introduced in
Refs. \onlinecite{parquet1} and \onlinecite{parquet2} for a
one-component Bose liquid, and for finding a local approximation for
the chain diagrams for the purpose of summing ladder diagrams. We find
here -- as in Ref. \onlinecite{MixMonster} -- that the very same
concept applies also in a different situation, namely in
demonstrating the relationship between the quasiparticle interaction
in FHNC theory and in the RPA. We stress that this result is an {\it
observation\/} on how the static FHNC approximation and the G0W
approximation of the self-energy are related.  It does {\it not\/}
mean that the FHNC approximation, which has been derived from looking
at the static structure function, and which is capable of reproducing
the static structure function with very good accuracy, is also {\it
adequate\/} for the calculation of single-particle properties and
Fermi liquid effects.

\section{APPLICATIONS}

\subsection{Hydrodynamic Effective Mass}
\label{hydrmass}

The first task is the calculation of the {\it hydrodynamic effective
mass.\/} This quantity describes the interaction of a single impurity
atom with the $^4$He background.  Due to the high density of the
background, the G0W approximation recovers only about 80 percent of
the effect and the more complete treatment of the time--dependent pair
correlations described in Sec. \ref{single} improves this result
considerably.

\begin{figure}
\vspace{0.5truein}
\centerline{\epsfxsize=5truein\epsffile{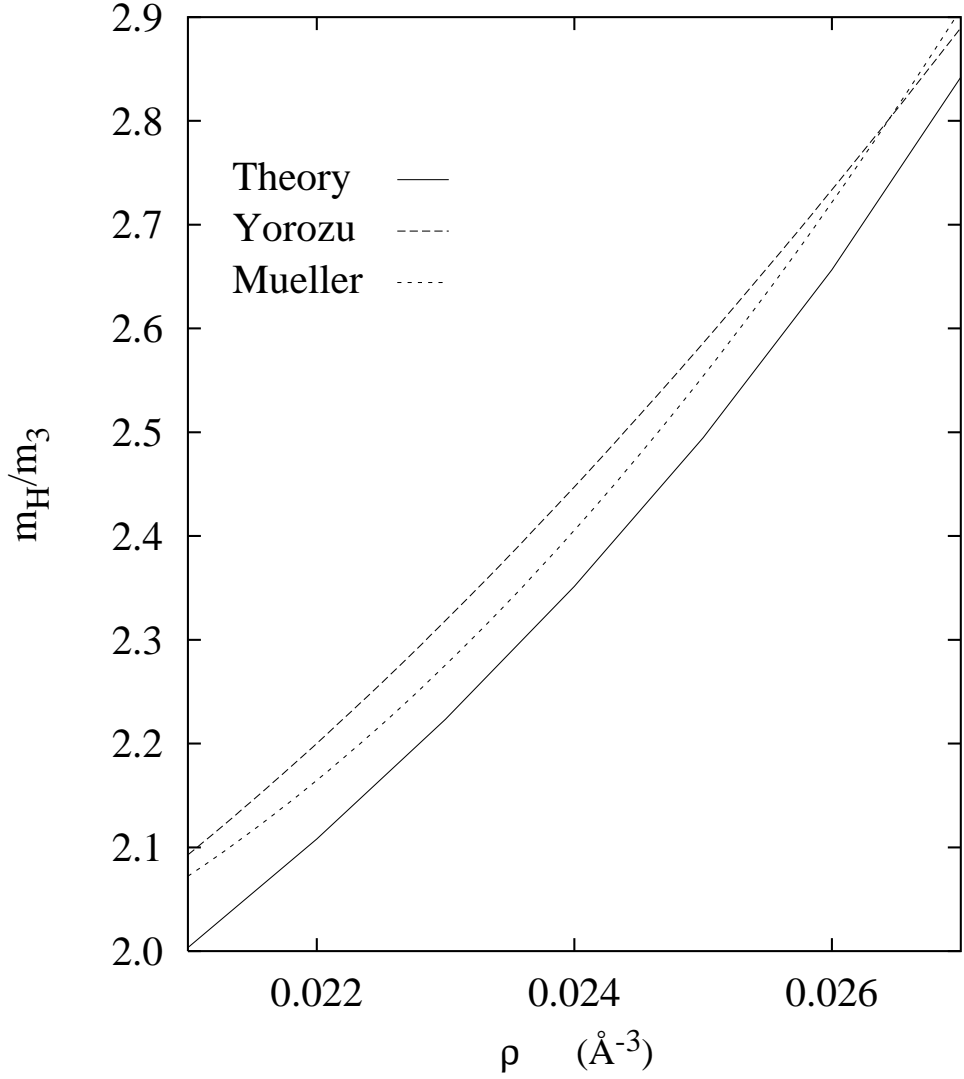}}
\vspace{0.5truein}
\caption{The figure shows the theoretical hydrodynamic mass from
our calculation described in Section \ref{single} (solid line), and
our zero-concentration extrapolations (\protect\ref{japmass}) and
(\protect\ref{julmass}) of the data of
Ref. \protect\onlinecite{Yorozu93} (long dashed line) and
\protect\onlinecite{SimonsLT21} (short dashed line).
\label{fig:effmass}}
\end{figure}

One of the formal results of our analysis is that the equations
(\ref{conII}), (\ref{self}) and (\ref{kernel}) treat the phonon-roton
spectrum as a Feynman-type spectrum,
$\epsilon^{(4)}(k)=\hbar^2k^2/[2m_4 S(k)]$, which is known to lie by a
factor of two too high in the roton region. To include the backflow
correction also into the motion of the $^4$He particles one should add
{\it fluctuating triplet correlations\/} into the wave function
(\ref{timedep}) and solve the triplet continuity equation. This leads
to a rather complicated integral equation with uncertainty in the
treatment of the {\it four-particle distribution function.\/} The
hierarchy of continuity equations must be truncated at some tractable
level. A reasonable shortcut is to modify the integral equation
(\ref{conII}) and the self energy (\ref{self}) by using the
experimental phonon-roton spectrum in $\epsilon^{(4)}(k)$, replacing
the impurity mass by its effective mass in the impurity kinetic energy
term, $t^{(3)}(k)$, and then ignoring the higher order
correlations.
For the triplet distribution function we have used the
convolution approximation together with the triplet correlation
function as described in the Appendix.\cite{MixMonster} The
self-consistent solution of Eq. (\ref{mhydro}) leads to a rather good
prediction of the hydrodynamic effective mass over the whole density
regime that is experimentally accessible. The results are shown in
Fig. \ref{fig:effmass} together with the fits to the measurements by
Yorozu {\it et al.} \cite{Yorozu93} and Simons and Mueller
\cite{SimonsLT21}. More details of those results are shown in Table
\ref{tab:hydmass}, which will be discussed in the next section.

\begin{figure}
\vspace{0.5truein}
\centerline{\epsfxsize=6.25truein\epsffile{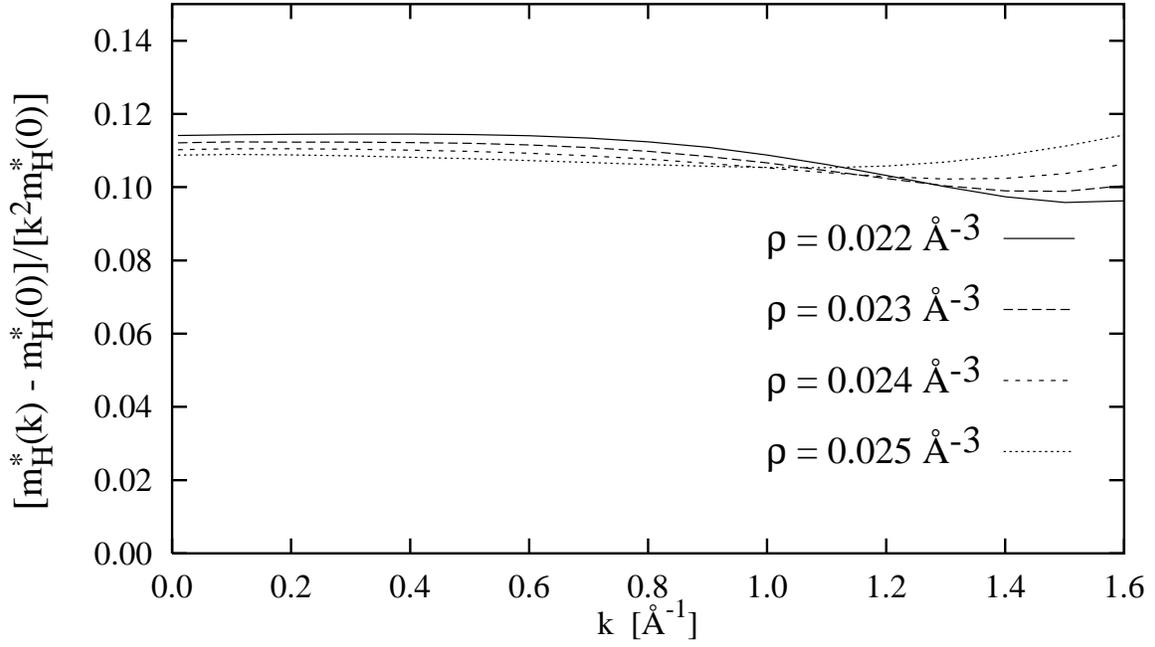}}
\vspace{0.5truein}
\caption{The {\it momentum dependent part\/} of the
hydrodynamic effective mass for densities between
0.022~${\rm\AA}^{-3}$ and 0.025~${\rm\AA}^{-3}$ as labeled in the
figure. To highlight the momentum--dependent part, we display,
according to Eq. (\protect\ref{kdep}), the
ratio $(m^*_H(k) - m^*_H(0))/(k^2m^*_H(0))$.
\label{fig:mhydro}}
\end{figure}

Next we need to determine the momentum dependence of the hydrodynamic
effective mass; it is necessary to study this effect because at finite
concentration the particles at the Fermi surface have a finite
momentum. For that purpose, we have calculated the dispersion relation
from Eq. (\ref{conI}) and have determined the ``momentum dependence''
of the hydrodynamic mass by writing the spectrum in the form
\begin{equation}
        \hbar\omega_H(k) = {\hbar^2 k^2\over 2m^*_H(k)}\ .
\end{equation}
In Fig. \ref{fig:mhydro} we plot the quantity
\begin{equation}
        {1\over k^2}\Bigl[{m^*_H(k)\over m^*_H(0)}-1\Bigr]\ ,
\label{kdep}
\end{equation}
which turns out to be {\it constant\/} for a wide momentum range
implying quadratic momentum dependence of the effective mass. Our
value of this constant is 0.114 at zero pressure. It has also been
estimated experimentally by F\aa k {\it et al.}.\cite{Fak} They find
the value 0.114$\pm$0.01 at the pressure P=0.1 atm and 1\%
concentration by fitting the momentum dependence of the spectrum in
the range 0.9\AA\,$^{-1}<k<1.65$\,\AA$^{-1}$. This is in very good
agreement with our result. The density dependence of this
constant is weak both in our calculations and in experimental
analysis. A slightly larger value 0.19 was obtained recently by
Fabrocini and Polls \cite{FabPollsSpect}. That is due to the fact that
their calculated spectrum is slightly softer and ours.

\begin{figure}
\vspace{0.5truein}
\centerline{\epsfxsize=6.25truein\epsffile{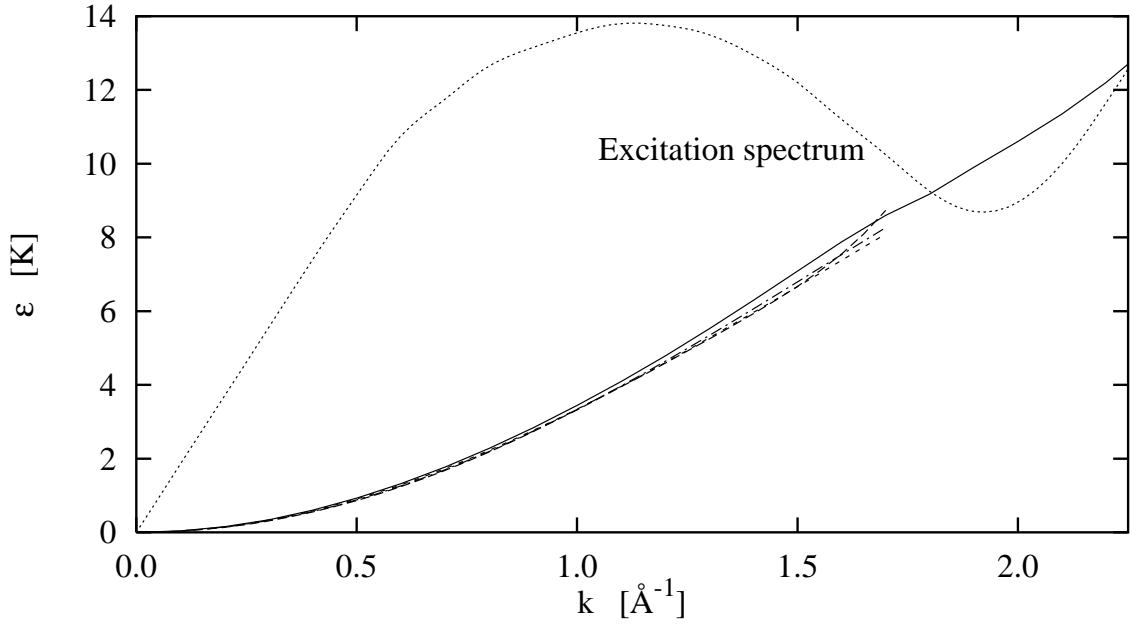}}
\vspace{0.5truein}
\caption{The excitation spectrum of the $^3$He impurity. The solid
curve is the result of the present theory. It is compared with the
measurements by Greywall (Ref. \protect\onlinecite{Greywall79}) (long dashed
line), F\aa k {\it et al.} (Ref. \protect\onlinecite{Fak}) ( short dashed line)
and Owers-Bradley {\it et al.} (Ref. \protect\onlinecite{OwersBra})
(dash-doted line). The dotted line shows, for reference, the experimental
phonon--roton spectrum.\protect\cite{CowleyWoods}
\label{fig:speche}}
\end{figure}

Different parametrizations of the experimental spectra for
$k\leq1.7$\,\AA\ have been suggested. \cite{Greywall79,OwersBra}
Fig. \ref{fig:speche} shows a comparison between our dispersion
relation and these parametrizations. Our spectrum is only slightly
stiffer, but the agreement appears to be satisfactory for all
practical purposes.

\subsection{Damping of Impurity Motion}
\label{spectrum}

At higher momenta and associated energy, it
becomes kinematically possible that the $^3$He decays into a roton and
a lower energy impurity mode. At this point, the self energy becomes
complex and no solution for the pole equation (\ref{conI}) can be
found.\cite{JacksonSumrules} The dynamic structure function has no
longer a delta function peak; instead the energy is distributed over a
range of modes. Therefore for momenta $k\geq1.7$\AA\ we plot in
Fig. \ref{fig:speche} the maximum of the dynamic structure function,
which shows that the impurity mode crosses the phonon-roton spectrum
smoothly. The peak broadens substantially before the second
crossing at $k \approx 2.25$\AA.

\begin{figure}
\vspace{0.5truein}
\centerline{\epsfxsize=6.25truein\epsffile{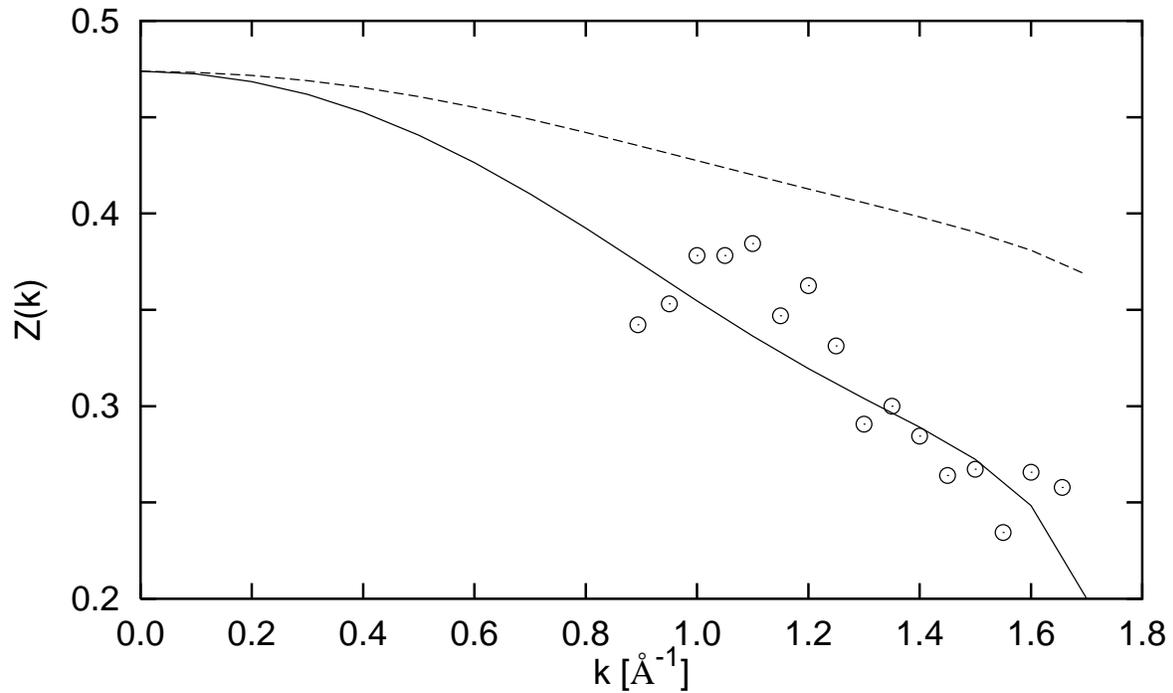}}
\vspace{0.5truein}
\caption{The pole strength of the elementary impurity excitation mode
is plotted as a function of momentum (solid line). The measured
strength of the particle-hole excitation at 1\% concentration and
saturation vapor pressure from Fig. 11 of
Ref. \protect\onlinecite{Fak} is shown with circles. For comparison we
also show the effective mass as a function of momentum.
\label{fig:zimpu}}
\end{figure}

The pole strength defined in Eq.(\ref{strength}) is calculated by
numerically differentiating the self energy. That is very reliable
because the self energy is a smooth function of $\omega$ in that
frequency range. The results are plotted in Fig. \ref{fig:zimpu}. As
pointed out in Eq.(\ref{zzero}) its long wavelength limit is inversely
proportional to the effective mass. The monotonically decreasing
behavior with decreasing wavelength is partly due to the increase of
the effective mass, also shown in the figure, and partly due to the
increasing contribution of the phonon-impurity channels at higher
frequences. F\aa k {\it et al.} ~\cite{Fak} have measured the area
under the particle-hole peak of the 1\% concentration mixture for the
momentum range 0.9\AA$^{-1}<k<$1.65\AA$^{-1}$. These results shown in
Fig.\ref{fig:zimpu} are in very good agreement with our
calculations. The measurements with 5\% concentration suggest only a
weak concentration dependence and thus the comparison with the zero
concentration limit calculations is justified.

\begin{figure}
\vspace{0.5truein}
\centerline{\epsfxsize=6.25truein \epsffile{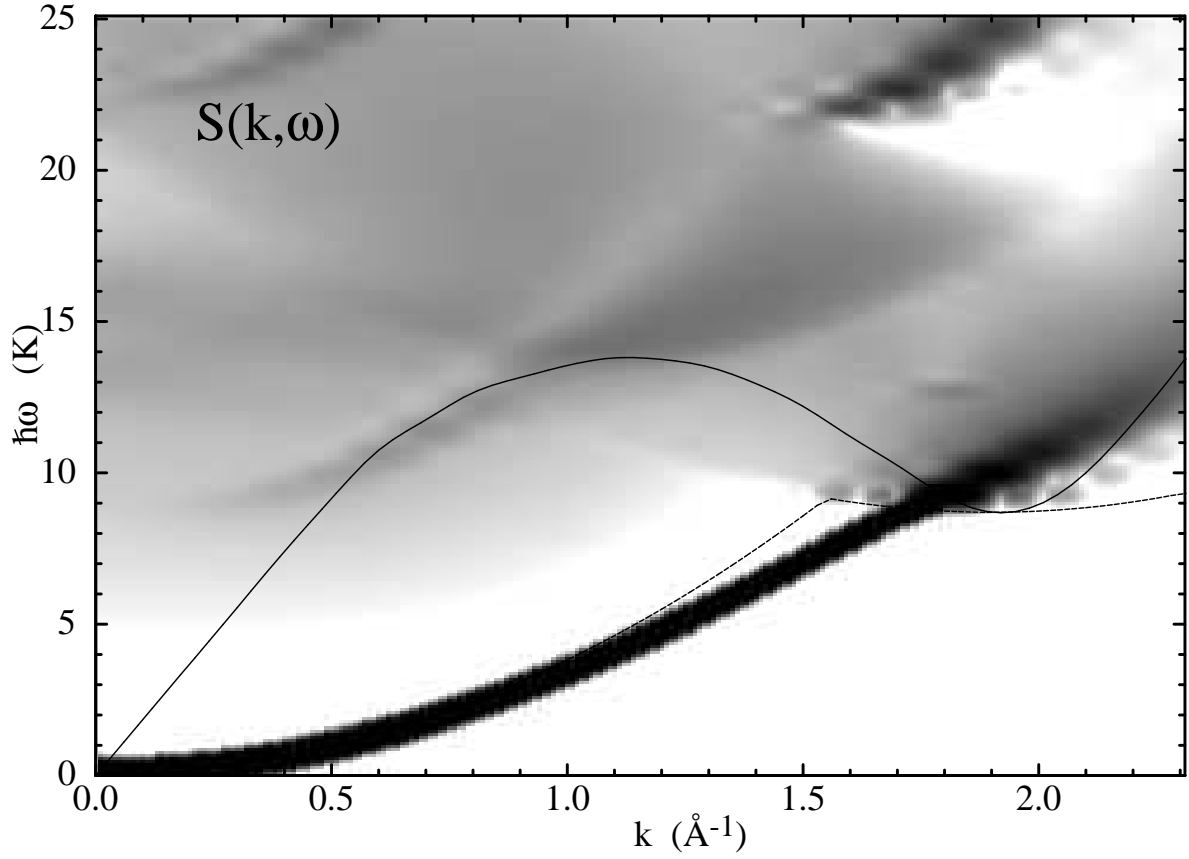}}
\vspace{0.5truein}
\caption{The impurity dynamic structure function $S^{(3)}(k,\omega)$
plotted in the $k,\omega$ plane. Also shown are the phonon-roton
spectrum of the background $^4$He (heavy solid line, the data are from
\protect\onlinecite{CowleyWoods}), and the decay threshold of the
impurity excitation mode (dashed line).
\label{fig:Skwgrey}}
\end{figure}

In Fig. \ref{fig:Skwgrey} we show the grey scale plot of the low
energy part of the $S(k,\omega)$ for $\omega<25~K$. The pole
contribution of the elementary excitation mode has been artificially
broadened by convolution with Gaussians in such a way that the area
under the peak is equal to the strength of the pole. In the figure we
also show the experimental phonon-roton spectrum and the continuum
boundary. The elementary excitation mode crosses that boundary at
1.72\,\AA$^{-1}$ and the decay into a roton becomes possible.  The
dominant peak in $S(k,\omega)$ crosses the phonon-roton spectrum and
looses its strength to higher lying modes. The mode slightly above 20
K is clearly visible in the figure.

\begin{figure}
\vspace{0.5truein}
\centerline{\epsfxsize=6.25truein\epsffile{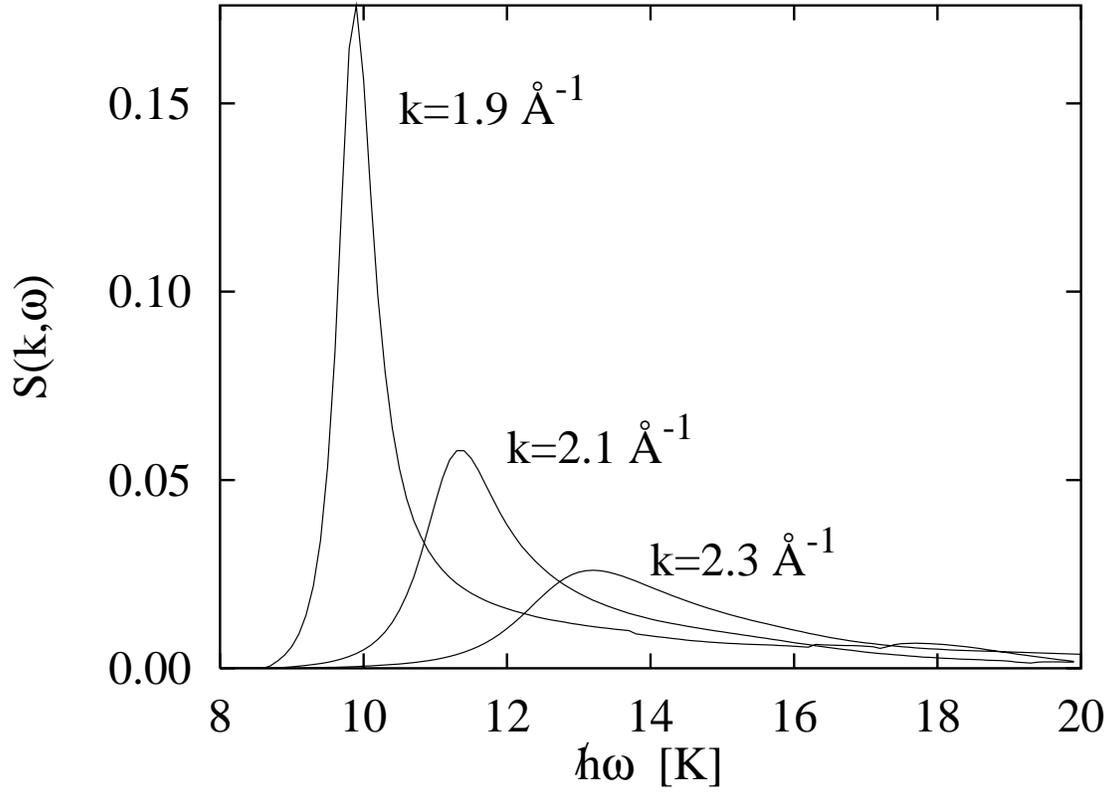}}
\vspace{0.5truein}
\caption{The dynamic structure function of the impurity
$S^{(3)}(k,\omega)$ in the roton region as a function of $\omega$ for
$k$=1.9\,\AA$^{-1}$, 2.1\,\AA$^{-1}$ and 2.3\,\AA$^{-1}$.
\label{fig:Skwcurves}}
\end{figure}

Damping of the impurity motion in the roton region is more clearly
shown in Fig. \ref{fig:Skwcurves}. There we plot $S(k,\omega)$ as a
function of $\omega$ for some typical values of $k$ = 1.9, 2.1 and 2.3
\AA$^{-1}$. The peak, which is a delta function when
$k<$1.72\,\AA$^{-1}$ broadens and looses its strength when the roton
minimum is past. In Table \ref{tab:penetra} we give the half widths of
those peaks. From them we can estimate the life time $\Delta$ of the
mode. The mode propagates with the group velocity and from that we can
calculate the distance $\lambda$ the impurity can travel before it
looses its energy to a roton,
\begin{equation}
        \lambda = {\hbar^2 k\over m_H^*(k) \Delta}\ .
\end{equation}
The propagation distance is typically a few \AA ngstr\"oms as 
also shown in Table \ref{tab:penetra}.

The penetration depth of a $^3$He atom at energies that are high enough
such that the impurity atom can couple to the roton are important
for the interpretation of recent measurements of the momentum transfer
of single $^3$He atoms from $^4$He clusters\cite{ToenniesPrivate}.
In these experiments, a significant increase of the momentum transfer has
been observed that is consistent with our energies. The very rapid
drop of the penetration depth shows that practically all $^3$He atoms
colliding with a $^4$He cluster will be absorbed from clusters
as small as 100 atoms, which have a diameter\cite{HNCdrops} of
about 25\,\AA. One of the reasons for this is the comparison of the
cluster size with the penetration depth, the other is that the excitation
spectrum of even such small clusters is remarkably similar to that
of the bulk liquid. The argument does not yet include the inelastic coupling
of the impurity atom to the lower--lying surface excitations, which can be
sizable\cite{scatt3}.

\subsection{Fermi Liquid Effective Mass}

We now turn to our microscopic calculation of Fermi liquid
effects. The necessary ingredients of the theory --- the effective
interactions $\tilde V_{\rm p-h}^{(\alpha\beta)}(q)$ and $\tilde W_{\rm
eff}(q)$ as well as the Feynman spectrum $\epsilon^{(4)}(k)$ --- have
been obtained in Ref. \onlinecite{MixMonster}. There are two sets of
accurate experimental data, those of Yorozu {\it et
al.}\cite{Yorozu93} and those of Simons and
Mueller.\cite{SimonsLT21} These experiments differ by the pressure
normalization, but lead otherwise to very similar results.

Clearly, the effective mass is {\it dominated\/} by the hydrodynamic
backflow as calculated above. To eliminate any uncertainty caused by
inaccuracies in that calculation, we have, for the further
calculations, {\it not\/} used the theoretical values obtained in
Section \ref{single} and shown in Figs. \ref{fig:mhydro} and
\ref{fig:speche}, but rather considered the zero-concentration limit of
the hydrodynamic mass as a phenomenological input. After the
concentration dependence was calculated from the Fermi liquid
contributions as described below, we have fitted our theoretical
values to the experiments of Refs. \onlinecite{Yorozu93} and
\onlinecite{SimonsLT21} to optimize the {\it overall\/} agreement and
then calculated the zero-concentration extrapolation.

\begin{figure}
\centerline{\epsfxsize=6.25truein\epsffile{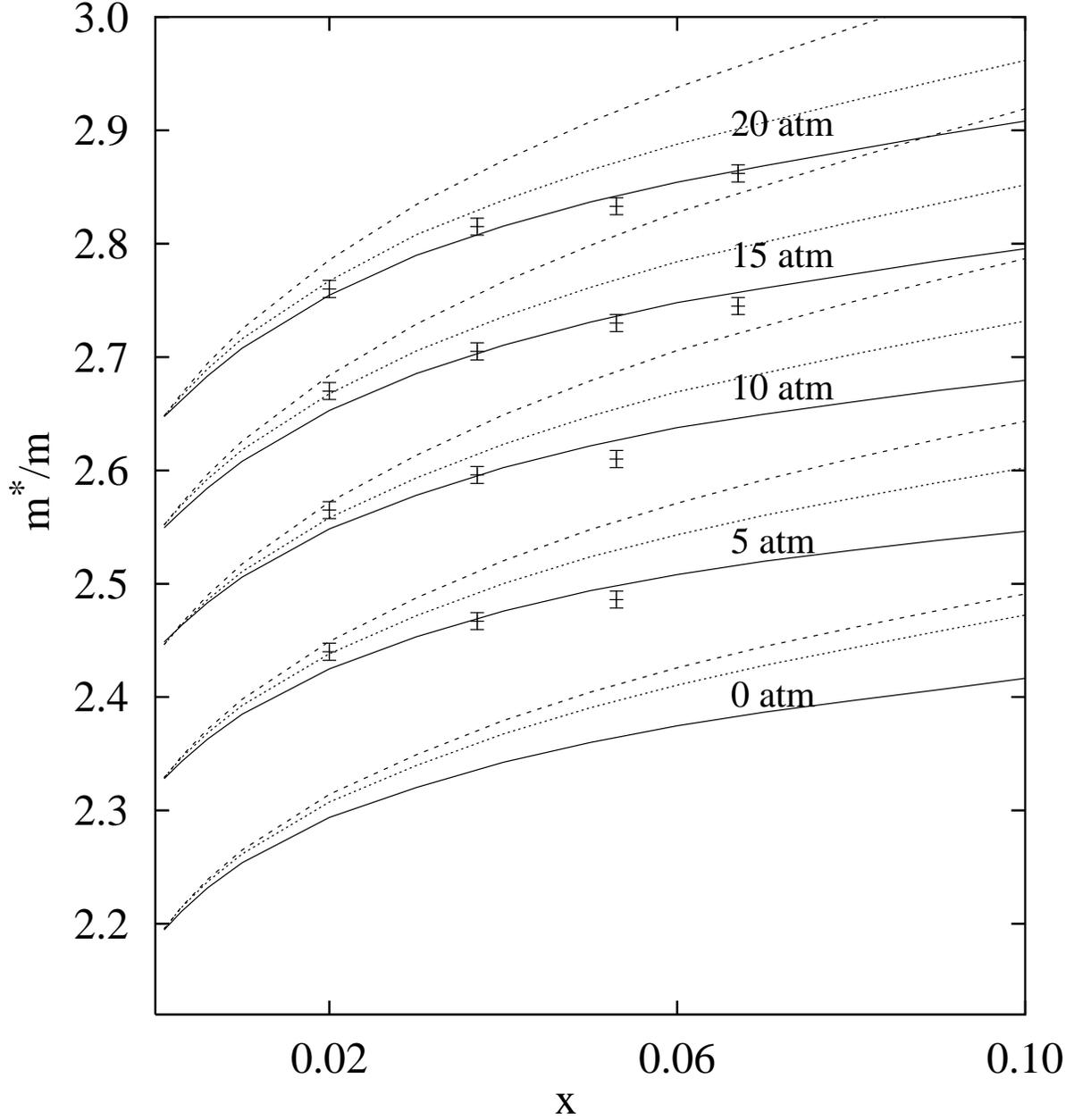}}
\vspace{0.5truein}
\caption{Theoretical and experimental effective mass ratio
$m^*(P,x)\over m$ as a function of pressure $P$ and concentration
$x$. The full curve is the fully self-consistent result, the
dotted curve is the result without retardation effects, {\it i.e.\/}
using the same quasiparticle interaction as in the spin--channel.
The short dashed
curve the static approximation. Symbols with error bars refer to the
data of Ref. \protect\onlinecite{Yorozu93}.
\label{fig:massj}}
\end{figure}

\begin{figure}
\centerline{\epsfxsize=6.25truein\epsffile{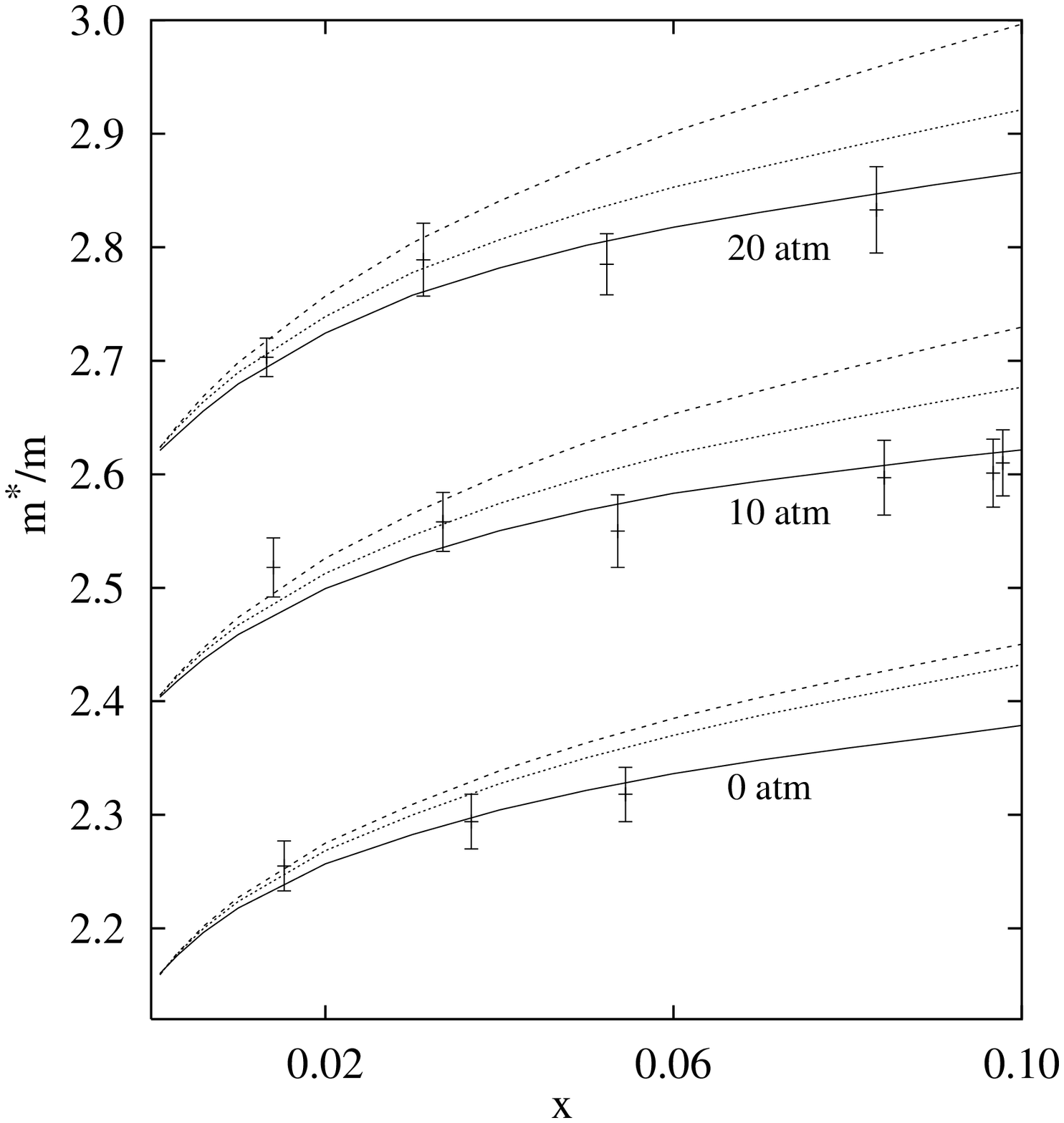}}
\vspace{0.5truein}
\caption{Same as Fig. \protect\ref{fig:massj} for the experiments
of Ref. \protect\onlinecite{SimonsLT21}.
\label{fig:massx}}
\end{figure}

Three calculations have been carried out to determine the Fermi liquid
contributions to the effective mass of the $^3$He component as a
function of concentration and pressure. The first calculation applied
the simple FHNC/EL theory and the static effective interaction
(\ref{pFHNC}). To account for the hydrodynamic backflow, one must
supplement the Fermion contribution (\ref{epsHF}) by the hydrodynamic
contribution $\hbar\omega_H(k)$; then the spectrum has the form
\begin{equation}
\epsilon^{(3)}(k) = \hbar\omega_H(k) + u(k) + U_0 \; ,
\label{epsHNC}
\end{equation}
where the Fermi correction $u(k)$ is given in Eqs. (\ref{uFHNC}),
(\ref{pFHNC}). When treated this way, the concentration dependence of
the effective mass derived from the spectrum (\ref{epsHNC}) is visibly
steeper than the experimental one, as seen in Figs.  \ref{fig:massj} and
\ref{fig:massx}.

In the next step, we calculate the effective mass using $\tilde V_{\rm
eff\/}({\bf k} ,0)$ as quasiparticle--interaction equivalently,
setting $\bar\omega = 0$ in Eq. (\ref{equiv}). This form of the
self-energy relaxes the approximations made by the FHNC theory since
it takes the effective interaction at the Fermi surface and not at an
average energy. We see in Figs. \ref{fig:massj} and \ref{fig:massx}
that the agreement with the experiment is indeed improved; the
approximation recovers about half of the discrepancy between the FHNC
approximation and experiments.

Finally, we have carried out self-consistent calculations of the
effective mass. It is sufficient for that purpose to assume a
single-particle spectrum of the form $t^{(3)}(k) = \hbar^2 k^2/ 2m^*$
in the Green's function (\ref{green}) and, consequently, in
Eq. (\ref{SigmaFinal}); note that the hydrodynamic mass is included in
the Green's function. This effective mass is then determined
self-consistently by requiring that the spectrum $\epsilon^{(3)}(k)$
determined by
\begin{equation}
        \epsilon^{(3)}(k) = \hbar\omega_H(k) +
        \Sigma_F\left(k,{\hbar k^2\over 2m^*}\right)
\label{epsFULL}
\end{equation}
can be fitted by the same effective mass that has been used in the
self-energy. This calculation provides a very good agreement with the
experimental data as seen in Figs. \ref{fig:massj} and
\ref{fig:massx}.  The agreement is worst for the pressure 10 atm and
the data of Ref. \onlinecite{SimonsLT21}; but we note that there is a
non-monotonic behavior of the slope of the data as a function of
pressure, and it might be interesting to re-examine this pressure
regime experimentally.

Carrying out this self--consistent procedure, we arrived at the
following interpolation formulas for the hydrodynamic mass:
\begin{equation}
        \left.{m^*_H\over m_3}\right)_{\rm expt} =
        2.18 + 2.43\; r + 2.67\; r^2  - 1.17\; r^3
\label{japmass}
\end{equation}
for the data of Ref. \onlinecite{Yorozu93} and
\begin{equation}
        \left.{m^*_H\over m_3}\right)_{\rm expt} =
        2.15 + 2.16 r\; + 4.47\; r^2
\label{julmass}
\end{equation}
from those of Ref. \onlinecite{SimonsLT21}. Here, $r =
\rho^{(4)}/\rho_0-1$, $\rho^{(4)}$ is the $^4$He density and
$\rho_0$=0.02183\AA$^{-3}$ is its value at the saturation vapor
pressure. Typically, the discrepancy between the two different
extrapolations is 0.03. These extrapolated hydrodynamic masses are
shown, together with our theoretical calculation of
Sec. \ref{single}, in Fig. \ref{fig:effmass}.  The theoretical
values are throughout the full density regime about 0.05 below the
experiments which is quite satisfactory given the level of
approximations.

To produce Figs. \ref{fig:massj} and \ref{fig:massx} we have used --
as stated before -- the hydrodynamic mass given in
Eqs. (\ref{japmass}) and (\ref{julmass}).  Since the calculations were
done for fixed densities at each concentration and the experiments
were done for fixed pressure, we have used the experimental
pressure-density relation of Ref. \onlinecite{OuY} to make the
conversion. Our calculations predict, at low concentration, a visible
curvature of the effective mass as a function of concentration,
similar to the predictions of Bashkin and
Meyerovich.\cite{BashkinMeye} Hence, we are not convinced that linear
extrapolations are a legitimate means to determine the hydrodynamic
mass unless concentrations are significantly lower than those examined
in Ref. \onlinecite{SimonsLT21}. Such a curvature is implicit to the
Fermi functions. Already the simple approximation (\ref{epsHNC}) would
lead to a behavior
\begin{equation}
m^*(x) = m^*_H +a x^{2/3} + b x + c x^{5/2} + d x^{7/3}\ldots\;.
\label{massfit}
\end{equation}
The numerical coefficients $a~\ldots~d$ can be calculated from the
moments of the potential, but such an expansion provides valid results
only for very small concentrations and a global fit of the form
(\ref{massfit}) to the calculated data is more accurate. In Table
\ref{tab:massfit} we list their values for different pressures obtained
from the least square fit to the fully self-consistent solution of
Eq. (\ref{epsFULL}).

On the other hand, we see no {\it drop\/} in the effective mass at
stable concentrations as, for example, proposed by Hsu and
Pines\cite{HsuPines}. The monotonic behavior is conclusively shown
experimentally by Simons and Mueller.\cite{SimonsLT21} The slight
downward curvature of the effective mass indicates that such a drop
will happen eventually; again the same conclusion can be drawn from
the analytic form (\ref{epsHNC}), but it does not happen at the
experimentally accessible stable concentrations.

\subsection{Magnetic Susceptibility}

We base our analysis on the susceptibility measurements by Ahonen {\it
et al.}.\cite{APCT76}  The spin susceptibility of the $^3$He component
is determined by both the effective mass and the quasiparticle
interaction in the spin channel,\cite{BBP66,BBP} {\it cf.\/}
Eq. (\ref{SuszLandau}), see also Ref. \onlinecite{BashkinMeye} for
further discussion. To extract the antisymmetric Landau parameter
$F_0^{\rm a}$ from the spin susceptibility, the effective mass must
therefore be known from an independent measurement. Again, one finds
that the spin susceptibility is vastly dominated by the one of the
free Fermi gas of particles with an effective mass $m^*_H$, and one
needs very accurate measurements to extract information on the
quasiparticle interaction. Ref. \onlinecite{APCT76} has used the best
values for the effective mass available at that time\cite{MassofAhon}
to calculate $F_0^{\rm a}$ from Eq. (\ref{SuszLandau}).

\begin{figure}
\centerline{\epsfxsize=6.25truein\epsffile{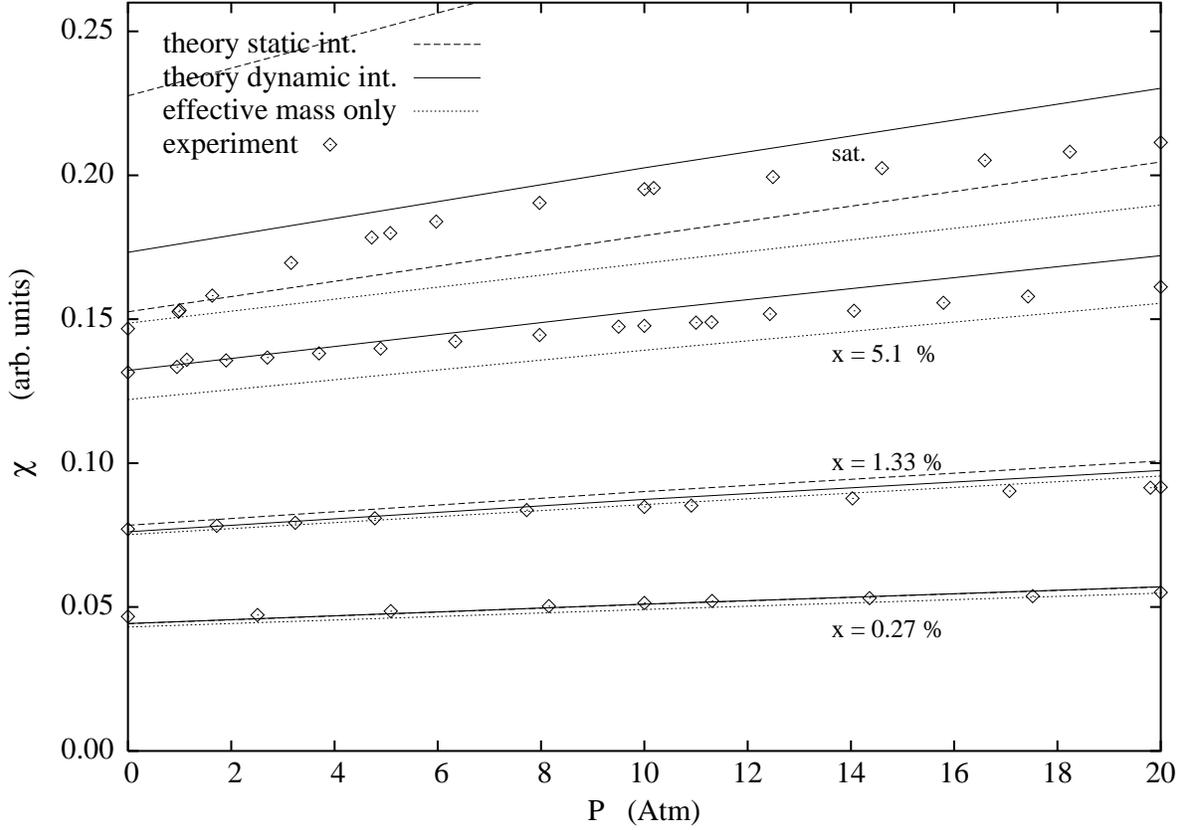}}
\vspace{0.5truein}
\caption{The figure shows a comparison between experimental
(diamonds, from Fig. 1 of Ref. \protect\onlinecite{APCT76}) and
theoretical magnetic susceptibilities. The vertical scale is arbitrary
and taken from Ref. \protect\onlinecite{APCT76}.  The theoretical
results were scaled to generate the best overall fit to the data at
$0.27\,\%$ and $1.33\,\%$ concentration. The solid lines are the
results of the full microscopic calculation, the long dashed lines the
results from using the static interaction $\tilde W_{\rm eff}(q)$.
Note that the static results for $5.1\,\%$ concentration are close
to the experiments at saturation concentration, and the static results
for $8.8\,\%$ percent are off-scale.
\label{fig:compchi}}
\end{figure}

To turn to the theoretical description, let us first look at the full
magnetic susceptibility obtained from Eq. (\ref{SuszLandau}) and the
effective interactions determined from our microscopic theory.
Parallel to the effective mass calculation, we use both the static
effective potential $\tilde W_{\rm eff}(q)$ and the RPA effective
interaction $\tilde V_{\rm eff}(q,0)$ to calculate the spin
susceptibility. The experimental data were given on an arbitrary
scale, we have therefore scaled each of our sets of results with a
global factor to provide the best overall fit to the data for the
concentrations 0.27\% and 1.33\%; the results are shown in Fig.
\ref{fig:compchi}. For comparison, we also show the susceptibilities that one
obtains from a free Fermi gas approximation with the effective mass
calculated in the preceding section. Up to five percent concentration,
the results look satisfactory; the theoretical results from the
dynamic theory and those from the free Fermi gas approximation
basically bracket the experimental data. There are some deviations at
the phase separation concentration; here the free Fermi gas model is
somewhat worse than the dynamic interaction. These deviations can, on
the theoretical side, be caused by the fact that the convergence of
our FHNC approximations becomes worse at higher Fermion
concentration. Part of the disagreement is also because the data at
lower pressures are given the phase-separation concentration and only
those at higher pressures at 8.8\,\%.  The limiting solubility at zero
pressure is approximately $x=0.065$~\cite{APCT76} and increases until it
reaches a maximum of about $x=0.095$ near $10 \;$ atm. The theoretical
calculations were, on the other hand, all done at 8.8\,\%
concentration.

A remarkable result is that the susceptibilities derived from the
static interaction $\tilde W_{\rm eff}(q)$ are clearly poor. In fact,
the independent particle approximation provides better agreement with
experiments than the static variational theory.  This is not entirely
unexpected in view of the above analysis of the effective mass, and
also in view of the much more severe deficiencies of the
Jastrow-Feenberg wave function for magnetic properties of pure
$^3$He. \cite{KSCP81}

\begin{figure}
\centerline{\epsfxsize=6.25truein\epsffile{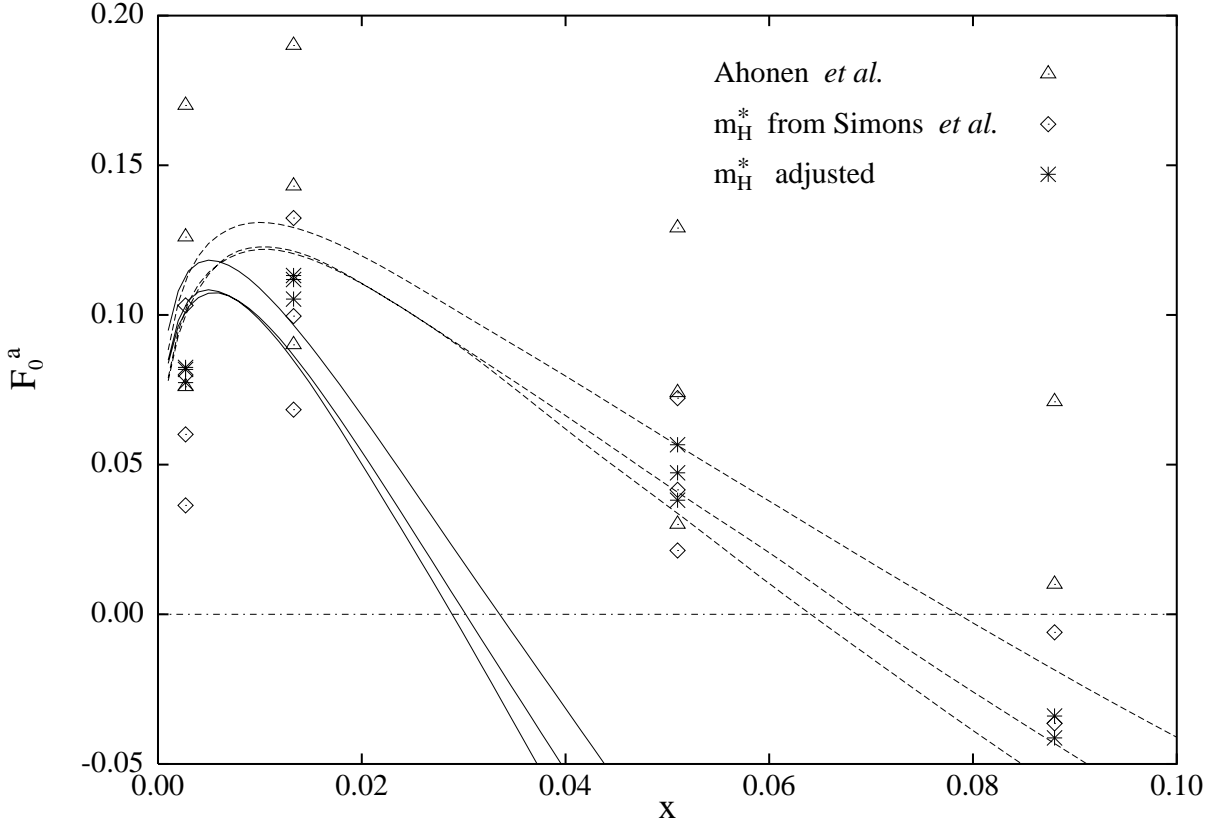}}
\vspace{0.5truein}
\caption{ The figure shows the calculated Landau
parameter $F_0^{\rm a}$ determined from the microscopic interactions
$\tilde V_{\rm eff}(q,0)$ (solid lines) and $\tilde W_{\rm eff}(q)$
(dashed lines) together with the results of Ahonen {\it et al.\/}
(boxes). The highest theoretical curves correspond to zero pressure
and the lowest to 20 atm. Also shown are estimates for the
Landau parameter $F_0^{\rm a}$ from the data of
Ref. \protect\onlinecite{APCT76}, using the effective masses from
Ref. \protect\onlinecite{Yorozu93} (diamonds). The lowest set of
symbols corresponds to saturation pressure, the medium set to $p =
10~$atm, and the upper set to $p = 20~$atm. There is no
experimental value at zero pressure and 8.8\% concentration.
\label{fig:F0as}}
\end{figure}

The purpose of the points made above was to raise a general awareness
of the difficulty of obtaining information {\it beyond\/} the
effective mass from susceptibility data in mixtures; this difficulty
has already been pointed out by the Ahonen {\it et al.\/}; further
concerns on the precision of the pressure normalization of that work
were raised by Rodriges {\it et al.\/}~\cite{RoVe97}. To extract the Landau
parameter $F_0^{\rm a}$ from these data, we have followed the
procedure of Ahonen {\it et al.\/}, but also used the more recent
effective masses data of Refs. \onlinecite{Yorozu93} and
\onlinecite{SimonsLT21} discussed above.  The resulting Landau
parameter $F_0^{\rm a}$ is shown, together with the original data of
Ref. \onlinecite{APCT76}, those obtained from the raw data with the
effective mass fit (\ref{julmass}), and our theoretical results, in
Fig. \ref{fig:F0as}. Obviously, the conclusions one can draw from the
same measurements depend visibly on what effective mass is used to
extract $F_0^{\rm a}$ from the spin--susceptibility. The agreement
between theory and experiment is significantly improved by using the
more recent effective mass data. There is still some vertical offset
but the {\it concentration dependence\/} of our theoretical results is
in fact quite good; the theory predicts a somewhat stronger
concentration dependence of $F_0^{\rm a}$ than is seen
experimentally. Most of the remaining disagreement can be removed by
using a slightly different hydrodynamic effective mass. To demonstrate
this, we have repeated the procedure of the preceding section and {\it
fitted\/} the hydrodynamic effective mass such that, at finite
concentrations, the theoretical Landau parameter is reproduced.  The
best fit to the data is reached by choosing the hydrodynamic effective
mass ratios given in the last column of Table \ref{tab:hydmass}.  It
appears that, apart from a slightly larger curvature suggested by
theory, there is little room for improvement. It is clear that any
attempt to extrapolate an effective mass from available susceptibility
data is uncertain because there are simply not enough data available
to carry out such extrapolations with confidence.

In this connection we also would like to stress the rather strong
concentration dependence in the low--concentration regime. It
indicates that, for the purpose of extracting $F_0^{\rm a}$ from
susceptibility measurements, even a concentration of 0.27\% is far
from the dilute limit. Note that --- if one assumes no readjustment of
the hydrodynamic mass --- the concentration dependence of $F_0^{\rm
a}$ as $x\rightarrow 0$ must be assumed to be even stronger. Similar
to the effective mass, the concentration dependence of the
Landau parameter $F_0^{\rm a}$ is most clearly discussed by writing
Eq. (\ref{f0a}) in coordinate space:

\begin{equation}
F_0^{\rm a} = -{m^* k_F\over 2 \pi^2 \hbar^2}\;
\int \! d^3 r \; V_{\rm eff}(r,0) \; j_0^2(r k_F)\,.
\label{f0ar}
\end{equation}

One source of the concentration dependence of $F_0^{\rm a}$ is {\it
kinematic\/} dependence coming from the explicit appearance of the
Fermi wave number $k_F$ in Eq. (\ref{f0ar}), which is caused by the
Pauli principle, as well the effective mass ratio in front of the
integral. The second is the implicit dependence of the effective
interaction $V_{\rm eff}(r,0)$ on the concentration. We found,
however, that this dependence is negligible within the experimentally
accessible regime under consideration here, and that $V_{\rm
eff}(r,0)$ is quite well represented by the interaction between a {\it
single pair\/} $^3$He atoms within the host liquid. In other words,
the concentration dependence is almost entirely due to the kinematics
dictated by the Pauli principle.

The coordinate--space representation (\ref{f0ar}) suggests a
concentration expansion similar to Eq. (\ref{massfit}) for the
antisymmetric Landau parameter. Since Eq. (\ref{f0ar}) suggests a
natural factorization into an effective mass ratio and an interaction
term, we expand
\begin{equation}
{m^*_H(x)\over m} \; F_0^{\rm a} (x)
 = a \; x^{1/3} + b \; x  + c \; x^{5/3} + d \; x^{7/3} \; .
\label{f0afit}
\end{equation}
        
The density--dependent parameters entering this fit are given in Table
\ref{tab:F0afit}. From the form of our results shown in
Fig. \ref{fig:F0as} it appears that the first two coefficients should
suffice; inclusion of two more terms improves the fit slightly. In
principle, one can obtain the first two coefficients from moments of
the effective interaction.  In practice, however, only the $x^{1/3}$
term gives a faithful approximation of our results up to 1.33~\%
concentration.

\begin{figure}
\centerline{\epsfxsize=6.25truein\epsffile{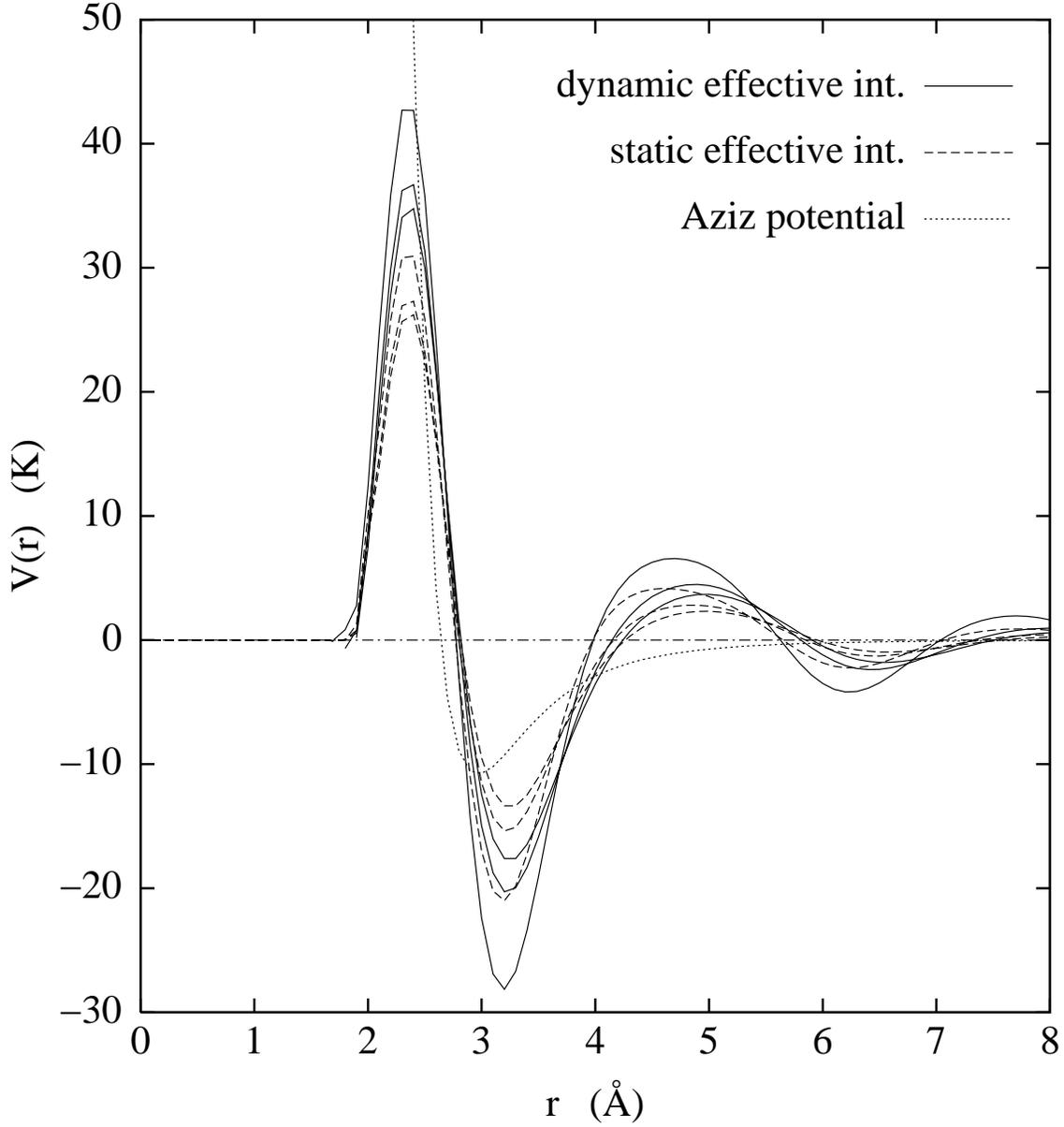}}
\vspace{0.5truein}
\caption{The figure shows the coordinate--space effective interactions
$V_{\rm eff}(r,0)$ (solid lines) and $W_{\rm eff}(r)$ (dashed lines)
for densities $\rho = 0.020~{\rm\AA}^{-3}$ below saturation, $\rho =
0.022~{\rm\AA}^{-3}$ close to saturation density, and $\rho =
0.026~{\rm\AA}^{-3}$ close to solidification. The potentials with the
higher repulsive peaks and deeper attractive wells correspond to the
higher pressure. The Aziz--potential
\protect\cite{Aziz} is shown for reference (dotted line).
\label{fig:veffofr}}
\end{figure}

Effective quasiparticle--interactions have in the past been discussed
and parameterized mostly in momentum space.\cite{EdwR,EbnE} This seems
to be appropriate because the quasiparticle interaction is indeed a
low-momentum property of the effective interaction. On the other hand,
the properties of these interactions are more intuitively described in
coordinate space. The effects that contribute to the effective
interaction have been discussed by Aldrich, Hsu, and
Pines;\cite{Aldrich,HsuPines} these are (a) core exclusion due to
short--range correlations, (b) an increase of the core--size due to
the kinetic energy needed to bend the relative wave function to zero
within the core, and (c) an increased attraction due to the presence
of other particles in the area where the potential is most
attractive. We show the effective interactions entering
Eq. (\ref{f0ar}) as a function of density at 5\,\% concentration in
Fig. \ref{fig:veffofr}; the underlying bare Aziz-potential\cite{Aziz}
is also shown for reference. All the effects predicted by Aldrich and
Pines are clearly seen in Fig. \ref{fig:veffofr}. We also note that
these overall features of the potential are quite resilient; the
density dependence of the interaction is relatively weak and the
concentration dependence practically invisible on the scale of Fig
\ref{fig:veffofr}. On the other hand, the differences between the
``dynamic'' and the ``static'' (FHNC) approximations are larger than
the variation of the potentials over the experimentally accessible
density regime of interest here.

Finally, a word is in order concerning the behavior of the dynamic
interaction within the core region. $V_{\rm eff}(r,0)$ should
be zero in this regime. In practice, it does not vanish vanish, this
is due to the ``RPA''-like approximation (\ref{effints}) for the
induced interaction. To be completely consistent, one would have to
solve the ring-- and ladder diagrams self--consistently using the full
Fermion propagators, and not the ``collective'' approximation. This
task has not been accomplished yet. We have therefore set the
interaction $V_{\rm eff}(r,0)$ to zero within the core region. Doing
so or not has no visible consequence, which lends credibility to our
CBF treatment that introduces the correct particle--hole propagators
{\it a posteriori\/} in a perturbative way.

\subsection{Scattering Matrix and Phase-Shifts}

To study the transport phenomena and potential superfluidity of the
$^3$He atoms in the medium, one needs to consider the scattering of
two $^3$He atoms at momenta ${\bf k}$ and ${\bf k}'$. Again, we must
consider both the hydrodynamic backflow, and the direct interaction
which is dominated, for long distances, by phonon exchange and for
short distances by the bare interaction between
particles.\cite{BashkinMeye,EbnE} A series of experiments to find a
superfluid phase transition of the $^3$He
component\cite{IFN89,Oh94,Pobell94} exhibit no evidence of
superfluidity down to a temperature of about $100~\mu{\rm K}$;
theoretical estimates\cite{Owen,Yoro92} of the transition temperature
range over several orders of magnitude.

Our calculations are similar to those of Owen,\cite{Owen} but use the
more accurate ground--state results of Ref. \onlinecite{MixMonster}.
The interaction entering the scattering equation is {\it not\/} the
same as the one used above for the calculation of Fermi liquid
parameters. The reason for this is that in the above calculation,
short--range correlations are included by calculating the
``parallel--connected'' diagrams or, in the language of perturbation
theory, the ``ladder diagrams''. In the scattering equation, these
short--range correlations are dealt with  by solving an effective
Schr\"odinger equation which reduces, for the case that both
scattering particles are in the many--body ground state, to the Euler
equation of the ground state theory. The scattering equation
is\cite{Owen}
\begin{equation}
\left[-{\hbar^2\over m_3}\nabla^2
+ V_{\rm scat}(r) - E\right]\phi({\bf r}) = 0\,,
\label{scateq}
\end{equation}
where
\begin{equation} 
V_{\rm scat}(r) =  V(r) +\Delta V_{\rm e}(r) + w_{\rm I}(r)\,.
\label{rspace}
\end{equation}
Here, $V(r)$ is the bare potential, $w_{\rm I}(r)$ the
``induced interaction'' of the FHNC theory, and
$\Delta V_{\rm e}(r)$ a correction due to elementary diagrams and
triplet correlations. In the {\it dilute limit\/}, the
pair-distribution function $g_{33}(r)$ between pairs of $^3$He atoms
is given by the zero-energy equation
\begin{equation}
\left[-{\hbar^2\over m_3}\nabla^2
+ V_{\rm scat}(r)\right]\sqrt{g_{33}({\bf r})} = 0\,.
\label{euler33}
\end{equation}
Eq. (\ref{euler33}) is one of the central equations of the HNC/FHNC
theory, see, for example. Ref. \onlinecite{MixMonster} for derivations
and discussion.

\begin{figure}
\centerline{\epsfxsize=6.25truein\epsffile{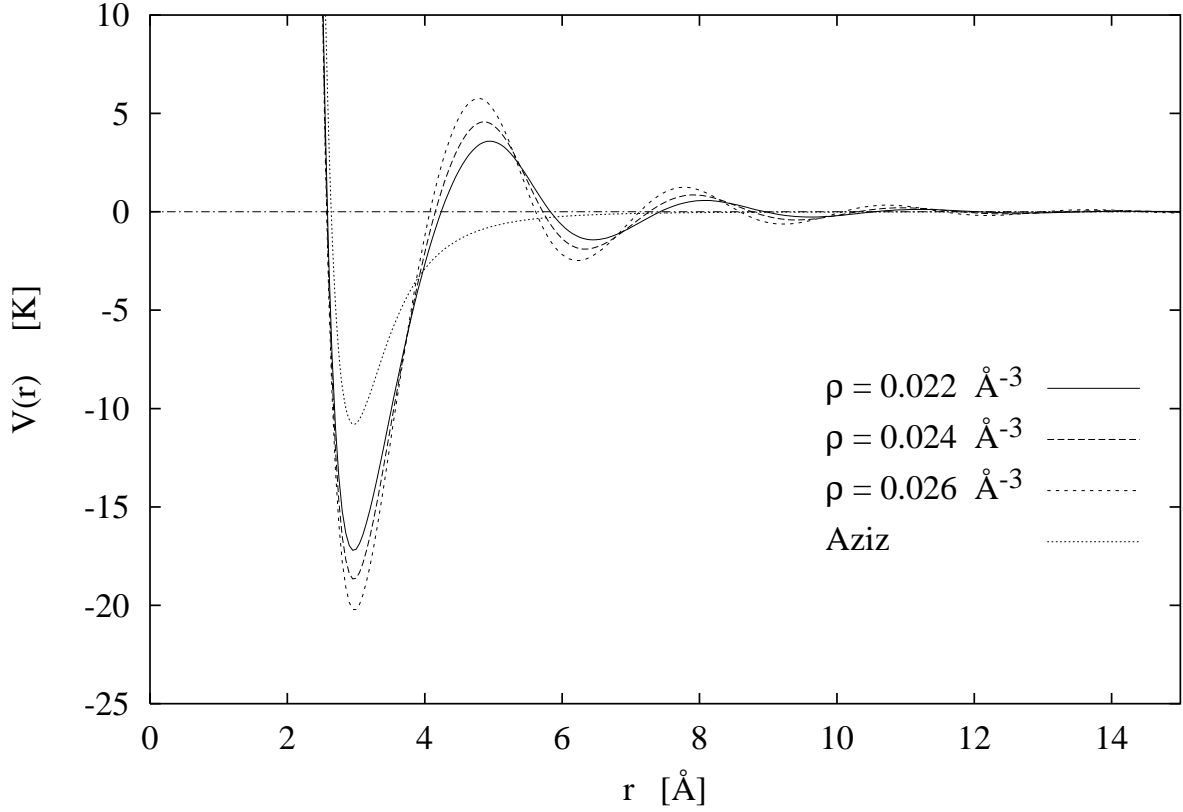}}
\vspace{0.5truein}
\caption{The figure shows the coordinate--space effective interactions
$V_{\rm scat}(r)$ entering the scattering equation (\ref{rspace})
at the density $\rho = 0.022\,{\rm\AA}^{-3}$ (solid line),
$\rho = 0.024\,{\rm\AA}^{-3}$ (long-dashed line) and
$\rho = 0.026\,{\rm\AA}^{-3}$ (short-dashed line).
The Aziz--potential
\protect\cite{Aziz} is shown for reference (dotted line).
\label{fig:vscatt}}
\end{figure}

The only difference to Owen's work is so far that we have included
elementary diagrams and triplet correlations in the ground state, and
hence the induced potential $w_{\rm I}(r)$ also changes. The effective
potential is shown, for three representative densities, in
Fig. \ref{fig:vscatt}. Again, this effective interaction is almost
independent of the $^3$He concentration; the quasiparticle scattering
potential depends therefore on the density and not on the
concentration. Although technically not necessary, it would be
acceptable to use (as Owen) the low-concentration limit of the
effective interaction to calculate scattering phase shifts.

One might now be led to argue that the $w_{\rm I}(r)$, which describes
phonon exchange, should also be a dynamic, energy--dependent
interaction as discussed above. This is in principle correct, but
unfortunately there is presently no practical way include such dynamic
effects. The reason for this is that the ground state equation
(\ref{euler33}) states that $\sqrt{g_{33}({\bf r})}$ is a {\it zero
energy\/} solution of the scattering equation (\ref{scateq}). Changing
the induced interaction {\it a posteriori\/} leads to an inconsistent
low-energy behavior of the solution and to spurious bound
states. Hence, a better calculation must await the developments of a
complete parquet-diagram theory that includes energy--dependent
interaction all the way through the diagram summations.

\begin{figure}
\centerline{\epsfxsize=6.25truein\epsffile{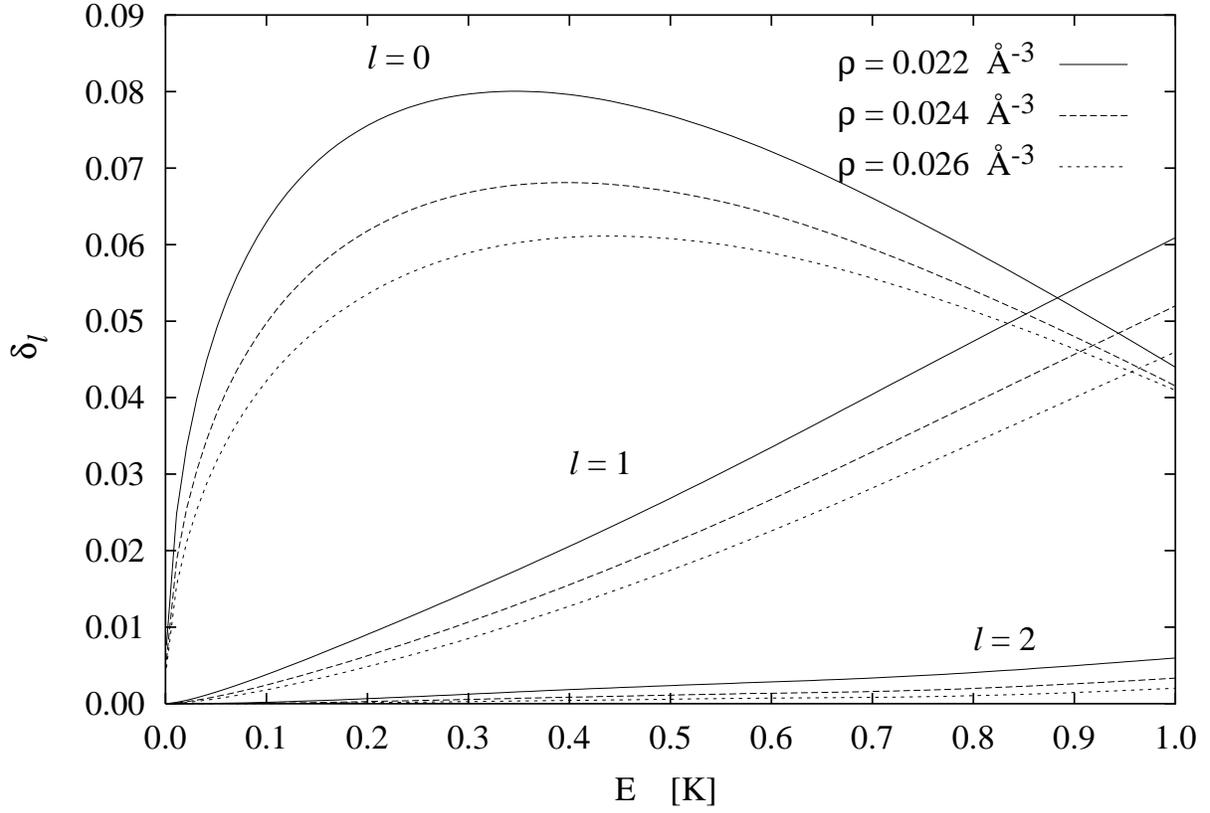}}
\vspace{0.5truein}
\caption{The $^3$He\,--$^3$He scattering phase--shifts are shown
at 1 percent concentration for the densities
$\rho = 0.022~{\rm\AA}^{-3}$ (solid line), $\rho =
0.024~{\rm\AA}^{-3}$ (long-dashed line), and  $\rho =
0.026~{\rm\AA}^{-3}$ (short-dashed line) for $\ell = 0,\, 1$, and $2$.
\label{fig:phases}}
\end{figure}

Returning to the original problem, we have solved the
scattering equation (\ref{scateq}) as a function of energy
in the $\ell = 0, 1$, and $\ell = 2$ channels. The corresponding
phase shifts $\delta_\ell$ are shown, as a function of density, in
Fig. \ref{fig:phases}.

The definition of the scattering amplitude is\cite{BaymPethick}
\begin{equation}
T_\ell(k) \; e^{-i\delta_\ell} = 
        - {4\pi \hbar^2 \over m_3 k} (2\ell + 1) \sin(\delta_\ell) \; ,
\end{equation}
where $k=\sqrt{2m_3 E}/\hbar$ is the relative momentum.

The most interesting application of our results is the estimate of a
potential superfluid phase transition. Manifestly microscopic
many--body theory is still at a rather unsatisfactory state when it
comes to predicting a superfluid phase transition in $^3$He, and the
mixture problem is no exception, possibly because of the lack of a
self--consistent parquet--like theory. At this point, we can only use
the scattering phase shifts in a weak--coupling approximation to
estimate the critical temperature of the phase transition.
\begin{equation}
\label{Tempc}
k_B T_c \approx  E_F
        \; \exp\left[{\pi^2 \, \hbar^2  \over m^* k_{\rm F} \;
      T_\ell(2 E_{\rm F})}\right] \; ,
\end{equation}
where $E_{\rm F} = {\hbar^2 \, k_{\rm F}^2 \over 2 \, m^*}$
is the Fermi energy and $T_\ell$ is evaluated at
$2 E_{\rm F}$ because two $^3$He with Fermi energy $E_F$
are forming a Cooper pair.  Eq. (\ref{Tempc})
gives only a rough estimate of the critical temperature
since it overestimates the critical temperature by two
orders of magnitude in pure $^3$He.

\begin{figure}
\centerline{\epsfxsize=6.25truein\epsffile{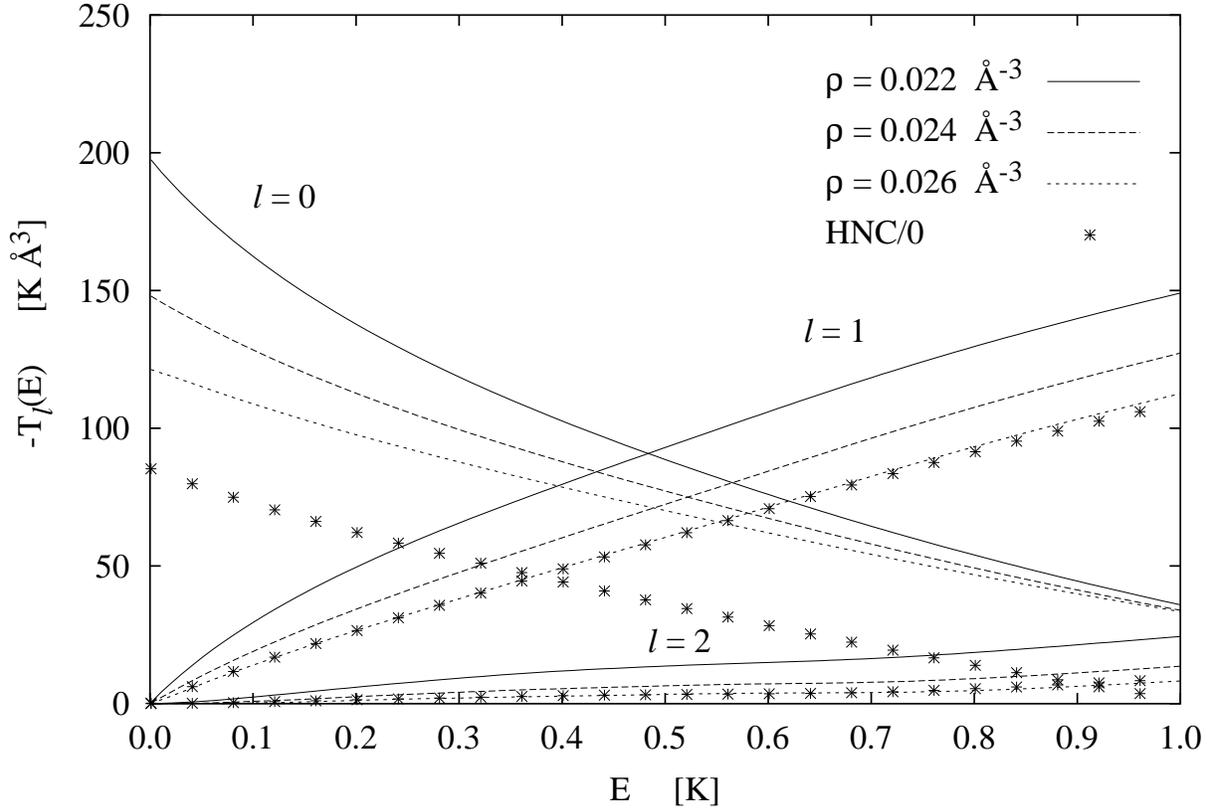}}
\vspace{0.5truein}
\caption{The $^3$He\,--$^3$He scattering amplitudes corresponding
to the phase shifts shown in Fig. (\ref{fig:phases})
at 1 percent concentration for the densities 
$\rho = 0.022~{\rm\AA}^{-3}$ (solid line),  $\rho =
0.024~{\rm\AA}^{-3}$ (long-dashed line), and  $\rho =
0.026~{\rm\AA}^{-3}$ (short-dashed line) for  $\ell = 0,\, 1$, and $2$.
The stars show the results in the simple HNC/0  approximation used 
by Owen \protect\cite{Owen}.
\label{fig:amplitudes}}
\end{figure}

\begin{figure}
\centerline{\epsfxsize=6.25truein\epsffile{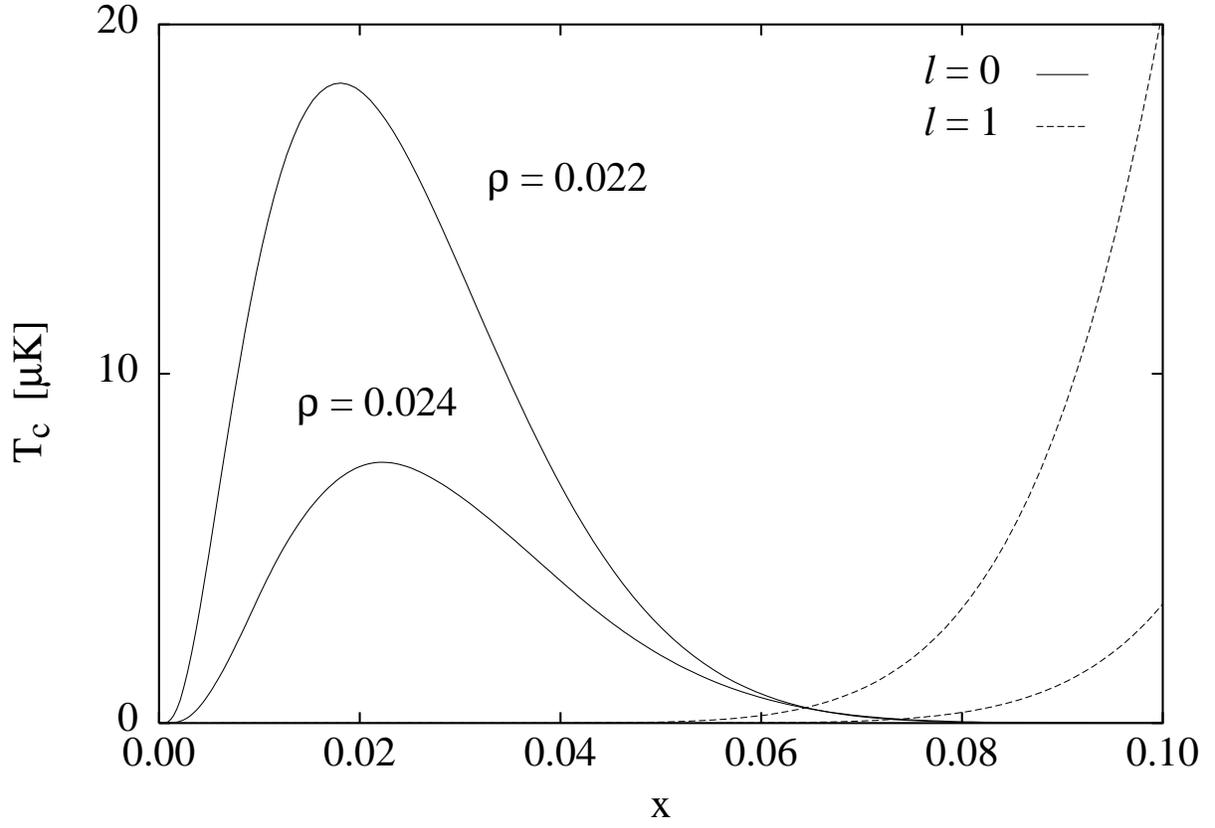}}
\vspace{0.5truein}
\caption{The transition temperatures for $s$--wave (solid lines)
and $p$--wave (dashed lines) pairing 
for the densities $\rho = 0.022~{\rm\AA}^{-3}$ (upper curves) and 
$\rho =0.024~{\rm\AA}^{-3}$ (lower curves). \label{fig:Transtemp}}
\end{figure}

Our results for the critical temperature are shown in 
Fig.~\ref{fig:Transtemp} for two representative densities.
The general trend is that the critical temperature
for $p$--wave pairing increases strongly with concentration, while
the temperature for the $s$--wave decreases; 
this is mainly due to the explicit appearance of the Fermi momentum 
in Eq. (\ref{Tempc}) and the resulting dependence of the scattering
matrix elements $T_\ell(E)$.
$T_c$ decreases with increasing pressure which can 
also be seen from the decrease of $T_0$ shown Fig.~\ref{fig:amplitudes}
as function of the energy.

Our results are an order of magnitude smaller than Owen's which we
have reproduced, but significantly larger than the values proposed in
Ref.~\onlinecite{Yoro92}. The difference to Owen's results is mainly
due to the smaller scattering matrix elements in both $s$ and $p$ wave
channels due to the more quantitative treatment of many--body
correlations.  However, the wide spread of results demonstrates
drastically the very sensitive dependence of the critical temperature
on the matrix elements. Therefore, while we believe
that the scenario shown in Fig.~\ref{fig:Transtemp} is qualitatively
accurate, we would like to be cautious about the quantitative
validity.

This is consistent with our calculation; however we
prefer to be cautious to call the results of our rather simple
calculation quantitative for reasons explained
above.

\section{SUMMARY}

We have in this paper analyzed various procedures for calculating the
quasiparticle interaction of the $^3$He component in a $^3$He\,--$^4$He
mixture. The technical parts of the calculation were based on the
equation of motions method, optimized (F)HNC theory of mixtures and on
CBF theory to infinite order. The close relationship between the
theories has been discussed.

Our results for the single--impurity spectrum are quite satisfactory.
One might object at this point that our use of the experimental
phonon--roton spectrum leads us away from manifestly microscopic
many--body theory, but we feel that such shortcuts are legitimate to
eliminate tedious and unrewarding computations whose outcome is
basically known.

Turning to the Fermion aspect of our calculations, we have seen that
the na\"\i ve FHNC theory is generally unsatisfactory for the
prediction of Fermi liquid effect compared with level of accuracy
that has been obtained for the microscopic calculation of ground state
properties of simple quantum liquids and quantum liquid mixtures. We
have also noted a systematic, stepwise improvement of the theory when
dynamic and Fermi surface specific effects are included.

The situation is clearest for the magnetic susceptibility where the
FHNC approximation takes the effective interaction at the energy
$\hbar\bar\omega(k)$ whereas it {\it should\/} have been taken at zero
energy. We stress that this is {\it not\/} a shortcoming of the (F)HNC
theory which is a specific method to sum large classes of diagrams. It
is a problem of the Jastrow-Feenberg function {\it per se.\/} It is
also clear how infinite order CBF theory resolves the problem in a
natural and elegant way, whereas it would be quite difficult to
describe the same physics by attempting to construct ``better''
variational wave functions for the ground state. The same effect is
also the explanation for both the miserable\cite{IndianMass}
performance of simple variational wave functions for the effective mass
in pure $^3$He, and for why magnetic properties of pure $^3$He are not
well described by the na\"\i ve application of variational wave
functions.  The power of the CBF approach lies in the possibility of
combining two generic many-body methods: Green's function concepts are
used for the examination of subtle, energy dependent effects and
variational methods for the unambiguous determination of static
effective interactions {\it whenever such interactions are
appropriate.}

In conclusion, it appears that one understand the {\it concentration
dependence\/} of the two Fermi liquid parameters $F_0^{\rm a}$ and
$F_1^{\rm s}$ reasonably well with a local quasiparticle interaction
if retardation effects are included. The remaining discrepancy between
our results for the effective mass and the magnetic susceptibility
seems to prevail at even very low concentrations where Fermi liquid
effects are particularly well described by our theory. The situation
becomes even more disturbing considering the fact that the data
at 0.27\,\% and 1.33\,\% concentration seem to indicate a smaller
curvature than theory, which should then turn steeper in order
to reach the independent particle limit $F_0^{\rm a}(x)\rightarrow 0$
as $x\rightarrow 0$. The simpler explanation is that the
extrapolations to zero concentrations are too uncertain. For more
accurate statements about the agreement (or disagreement) between
theory and experiment, one should have many more susceptibility
data especially at low concentrations.

The next step in the theoretical procedure would be the summation of
properly antisymmetrized exchange diagrams as suggested by Babu and
Brown.\cite{BabuBrown} But in view of the above discussion and the
very high accuracy that is needed to extract the first antisymmetric
Landau parameter from susceptibility data it might well be worthwhile
to reconsider these experiments.

\section*{ACKNOWLEDGMENTS}
We would like to dedicate this paper to the memory of our late
colleague Eugene Bashkin who has contributed much to the understanding
of $^3$He\,--$^4$He mixtures and in particular their magnetic properties.
The work was supported, in part, by the National Science Foundation
under grant DMR-9509743, the Austrian Science Fund (FWF) under project
P11098-PHY (to EK), and the Academy of Finland (to MS). Discussions
with C. E. Campbell, R. Mueller, M. Paalanen and A. Polls are gratefully
acknowledged. We thank A. Fabrocini and A. Polls
for giving us access to Ref. \onlinecite{FabPollsSpect} prior to
publication, as well as J. Harms and J. P. Toennies for informing us about 
their experiments\cite{ToenniesPrivate} well before publication.
 
\appendix
\section*{Derivation of the self energy}
In this Appendix we give the derivation of the self energy starting
from the continuity equations (\ref{conI}) and (\ref{conII}). The
first continuity equation defines the self energy in terms of the
solution of the second continuity equation. We explain our method for
solving these equations in momentum space leading to Eqs. (\ref{continuityI})
and (\ref{continuityII}) and use the notations of
Ref. \onlinecite{SKJLTP}.

The impurity pair- and triplet distribution functions are defined in
the usual way,
\begin{equation}
        g^{(34)}({\bf r}_0,{\bf r}_1) =
        {1\over\rho^{(3)}\rho^{(4)}} \;
        \rho^{(34)}({\bf r}_0,{\bf r}_1)\ ,
\label{g34}
\end{equation}
\begin{equation}
        g^{(344)}({\bf r}_0,{\bf r}_1,{\bf r}_2) =
        {1\over\rho^{(3)}\rho^{(4)} \rho^{(4)}}  \;
        \rho^{(344)}({\bf r}_0,{\bf r}_1,{\bf r}_2)\ ,
\label{g344}
\end{equation}
and the impurity structure function $S^{(34)}(k)$ is the Fourier transform
of the pair-distribution function $g^{(34)}(r)-1$. Analogous
definitions are used for the background $^4$He distribution and
structure functions.

Since our background $^4$He liquid is homogeneous the continuity
equations are easiest to solve in Fourier space. We have defined the
Fourier transform of the fluctuating one--particle density in
Eq. (\ref{deltarho}), for the time--dependent two--particle
correlation function it has the following form:
\begin{eqnarray}
        \alpha^{(34)}_{{\bf k},\omega}({\bf p})
        &=&\frac{\rho^{(3)}\rho^{(4)}}{\delta\rho^{(3)}(k,\omega)}
        \int \! d^3r_0\, d^3r_1 \, dt \; e^{-i({\bf k}\cdot{\bf r}_0+
        {\bf p}\cdot({\bf r}_0-{\bf r}_1)-\omega t)} \; 
        \delta u^{(34)} ({\bf r}_0,{\bf r}_1;t) \ .
\label{alpha}
\end{eqnarray}
Using these notations the first continuity equation
(\ref{continuityI}) transforms into the form 
\begin{equation}
        \hbar\omega-\frac{\hbar^2 k^2}{2m_3} - \Sigma^{(3)}(k,\omega)
        = {\tilde U_{\rm ext}(k,\omega)\over \delta\rho^{(3)}(k,\omega)} 
\label{conIk}
\end{equation}
with the self energy
\begin{equation}
        \Sigma^{(3)}(k,\omega)=\frac{\hbar^2}{2m_3} \int {d^3p\over
        (2\pi)^3\rho^{(4)}}~ {\bf k}\cdot{\bf p} \; S^{(34)}(p)\; 
        \alpha^{(34)}_{{\bf k},\omega} ({\bf p}) \, . 
\label{Aself}
\end{equation}

The second continuity equation (\ref{continuityII}) contains the
triplet distribution function. In order to calculate its Fourier
transform we define the triplet structure function
\begin{equation}
        S^{(344)}({\bf r}_0,{\bf r}_1,{\bf r}_2)=
        g^{(344)}({\bf r}_0,{\bf r}_1,{\bf r}_2)
        -g^{(34)}({\bf r}_0,{\bf r}_1)-g^{(34)}({\bf r}_0,{\bf r}_2)
        -g^{(44)}({\bf r}_1,{\bf r}_2)+2 \ .
\label{s344r}
\end{equation}
Our approximation for this quantity \cite{MixMonster,SKJLTP}
includes the convolution approximation together with the triplet
correlation function. In the momentum space that leads to the result

\begin{eqnarray}
        \tilde S^{(344)}({\bf k}_1,{\bf k}_2,{\bf k}_3)&=&
        \delta({\bf k}_1+{\bf k}_2+{\bf k}_3) \Big\{
        S^{(34)}({\bf k}_2)\; S^{(34)}({\bf k}_3)+S^{(34)}({\bf k}_1)
        \left[S^{(44)}({\bf k}_2)\; S^{(44)}({\bf k}_3)-1\right]
        \nonumber \\
        &+&\tilde u^{(344)}_3({\bf k}_1,{\bf k}_2,{\bf k}_3)
        \left[S^{(34)}({\bf k}_1)+1\right]\;
        S^{(44)}({\bf k}_2)\; S^{(44)}({\bf k}_3)\Big\}\ .
\label{s344k}
\end{eqnarray}
The second continuity equation (\ref{continuityII}) transforms in the
momentum space into the following integral equation
\begin{equation}
        [\hbar\omega-t^{(3)}({\bf k}+{\bf p})-\epsilon^{(4)}(p)]\; 
        \alpha^{(34)}_{{\bf k},\omega}({\bf p})
        =\hbar\omega\; \frac{{\bf k}\cdot{\bf p}}{k^2}\; 
        \frac{S^{(34)}(p)}{S^{(44)}(p)} 
        -\int {d^3q\over(2\pi)^3\, \rho^{(4)}}\; 
        \alpha^{(34)}_{{\bf k},\omega}({\bf q}) \;
        K_{k,\omega}({\bf p},{\bf q})\ , 
\label{intealpha}
\end{equation}
where the kernel $K_{k,\omega}({\bf p},{\bf q})$ is
\begin{eqnarray}
        K_{k,\omega}({\bf p},{\bf q})&=&
        S^{(44)}(q)\left\{\left[\left(S^{(34)}(|{\bf p}-{\bf q}|)+1\right)
        \tilde u^{(344)}_3({\bf p}-{\bf q},-{\bf p},{\bf q})
        +S^{(34)}(|{\bf p}-{\bf q}|)\right]
        \right. 
\nonumber\\
&\times& \left. 
        \left[\hbar\omega
        -\frac{({\bf k}+{\bf p})\cdot({\bf k}
        +{\bf q})}{p^2} \; t^{(3)}(p)\right]\right\}
        -S^{(34)}(|{\bf p}-{\bf q}|) \; \frac{{\bf p}\cdot{\bf q}}{p^2}\; 
        \epsilon^{(4)}(p) \ . 
\label{kernelA}
\end{eqnarray}

The singularity structure of the self energy as well as the integral
equation is best accounted by introducing the following notation,
\begin{equation}
        \beta^{(34)}_{{\bf k},\omega} ({\bf p})
        \equiv [\hbar\omega-t^{(3)}({\bf k}+{\bf p})-\epsilon^{(4)}(p)]\;
        \alpha^{(34)}_{{\bf k},\omega} ({\bf p})\ .
\label{betaA}
\end{equation}
Inserting this into Eqs. (\ref{Aself}) and (\ref{intealpha}) we get our
final forms for the self energy 
\begin{equation}
        \Sigma^{(3)} (k,\omega) =  {\hbar^2 \over 2\, m_3} 
        \int {d^3 p \over (2
        \pi)^3 \, \rho^{(4)}}\; {{\bf k}\cdot{\bf p} \;S^{(34)}(p)\;
        \beta^{(34)}_{{\bf k},\omega}({\bf p})
        \over\Bigl[\hbar\omega-t^{(3)}({\bf k}
        +{\bf p})-\epsilon^{(4)}(p)\Bigr]}\ ,
\end{equation}
and the integral equation
\begin{equation}
        \beta^{(34)}_{{\bf k},\omega}({\bf p})
        =\hbar\omega \;\frac{{\bf k}\cdot{\bf p}}{k^2}\;
        \frac{S^{(34)}(p)}{S^{(44)}(p)}
        -\int{d^3q\over (2\pi)^3 \,\rho^{(4)}} \;
        \frac{\beta^{(34)}_{{\bf k},\omega}({\bf q})\;
        K_{{\bf k},\omega}({\bf p},{\bf q})}
        {\Bigl[\hbar\omega-t^{(3)}({\bf k}+{\bf q})
        -\epsilon^{(4)}(q)\Bigr]}\ .
\end{equation}

The angle integration can be done conveniently by expanding
$\beta^{(34)}_{{\bf k},\omega}({\bf p})$ in terms of Legendre
polynomials,
\begin{equation}
        \beta^{(34)}_{{\bf k},\omega}({\bf p})
        =\sum_{l=0}^\infty \beta^{(34)}_l(k,\omega,p) P_l(x_{kp})\ . 
        \label{betakeh}
\end{equation}
Here $x_{kp}=\cos\theta_{kp}$ with the angle $\theta_{kp}$ between
vectors ${\bf k}$ and ${\bf p}$.
Similarly we expand the other angle dependent quantities
\begin{eqnarray}
        S^{(34)}(|{\bf p}-{\bf q}|)
        &\equiv&\sum_{l=0}^\infty s_l(p,q) P_l(x_{pq})  \; ,
\nonumber \\
        \frac{{\bf p}\cdot{\bf q}}{pq}\; S^{(34)}(|{\bf p}-{\bf q}|)
        &\equiv& \sum_{l=0}^\infty \sigma_l(p,q) P_l(x_{pq}) \; ,
\nonumber \\
        \left(S^{(34)}(|{\bf p}-{\bf q}|)+1\right)
        \tilde u^{(344)}_3({\bf p}-{\bf q},-{\bf p},{\bf q})
        &\equiv&\sum_{l=0}^\infty s_l'(p,q) P_l(x_{pq}) \; ,
\nonumber \\
        \frac{{\bf p}\cdot{\bf q}}{pq}
        \left(S^{(34)}(|{\bf p}-{\bf q}|)+1\right)
        \tilde u^{(344)}_3({\bf p}-{\bf q},-{\bf p},{\bf q})
        &\equiv&\sum_{l=0}^\infty \sigma_l '(p,q) P_l(x_{pq}) \ . 
\label{keh1}
\end{eqnarray}

Using these notations we can perform the angle integration
analytically in terms of Legendre polynomials and \emph{the complex
Legendre functions of the second kind} $Q_l(z)$,
\begin{equation}
        Q_l(z)={1\over2}\int_{-1}^{1} dx\,\frac{P_l(x)}{z-x} \, .
\end{equation}

That leads to the coupled integral equation for the expansion
coefficients $\beta^{(34)}_l(k,\omega,p)$
\begin{eqnarray}
        \beta^{(34)}_l(k,\omega,p)&+&\frac{1}{4\pi^2\rho^{(4)}}\int q^2dq\, 
        \sum_m \beta_m^{(34)}(k,\omega,q)
        \left\{ \frac{1}{kq}{\cal I}_{lm}^1(z_{kq})
        \left[S^{(44)}(q)[\bar{\hbar\omega}\bar s_l(p,q)
        \right. \right. 
\nonumber \\
        &-& \left.  \left. 
        k^2\bar s_l(p,q)-pq\bar\sigma_l(p,q)]
        -\frac{\bar\epsilon(p)}{p}q\sigma_l(p,q) \right] 
        -{\cal I}_{lm}^2(z_{kq})S^{(44)}(q)\bar s_l(p,q)
        \right.
\nonumber \\
        &-&\left. \frac{p}{q}{\cal I}_{lm}^3
        (z_{kq},p,q)(2l+1)S^{(44)}(q) \right\} 
        =\bar{\hbar\omega}\frac{p}{k}\frac{S^{(34)}(p)}{S^{(44)}(p)}
        \delta_{l,1}\ .
\label{legecom}
\end{eqnarray}
Energies are expressed in units which include the effective mass,
\begin{eqnarray}
        \bar{\hbar\omega} &=& {2 m^*_H\hbar\omega\over \hbar^2} \; , 
\nonumber\\
        \bar\epsilon(p)&=& {2 m^*_H\epsilon^{(4)}(p)\over \hbar^2}\ , 
\label{enegyunit}
\end{eqnarray}
and we have also introduced notations 
\begin{eqnarray}
        z_{kq}&=&\frac{\bar{\hbar\omega}-k^2-q^2-\bar \epsilon(q)}{2kq} 
\nonumber\\
        \bar s_l(p,q) &=& s_l(p,q)+s_l'(p,q) 
\nonumber\\
        \bar \sigma_l(p,q) &=& \sigma_l(p,q) + \sigma_l'(p,q) \ .
\label{notations}
\end{eqnarray}
The functions ${\cal I}_{lm}^1(z_{kq})$, ${\cal I}_{lm}^2(z_{kq})$ and
${\cal I}_{lm}^3(z_{kq},p,q)$ are the results of the angle integration
\begin{eqnarray}
{\cal I}_{lm}^1(z_{kq})&=&\left\{
\begin{array}{l}
P_m(z_{kq}) Q_l(z_{kq}),\, \mbox{when } m\leq l \\
Q_m(z_{kq}) P_l(z_{kq}),\, \mbox{when } m>l
\end{array} \right.\; , \\
\, \nonumber \\
{\cal I}_{lm}^2(z_{kq})&=&\left\{
\begin{array}{l}
z_{kq} P_m(z_{kq}) Q_l(z_{kq}),\, \mbox{when } m<l \\
z_{kq} Q_m(z_{kq}) P_l(z_{kq}),\, \mbox{when } m>l \\
z_{kq} P_l(z_{kq}) Q_l(z_{kq})-1/(2l+1),\, \mbox{when } m=l
\end{array} \right. \; ,\\
\, \nonumber \\
{\cal I}_{lm}^3(z_{kq},p,q)&=&\sum_n\left(
\begin{array}{ccc}
l&1&n \\
0&0&0
\end{array}
\right)^2 \bar s_n(p,q)
\left\{
\begin{array}{l}
P_m(z_{kq}) Q_n(z_{kq}),\, \mbox{when } m\leq n \\
Q_m(z_{kq}) P_n(z_{kq}),\, \mbox{when } m>n
\end{array} \right. \, ; 
\end{eqnarray}
with the $3j$-symbol $\left( \begin{array}{ccc} l&1&n \\ 0&0&0
\end{array} \right)$.

Using the above notations the self energy has the form
\begin{equation}
        \Sigma^{(3)} (k,\omega) = {m^*_H\over m_3}\int_0^\infty 
        {p^2 dp\over 4 \pi^2\rho^{(4)}} S^{(34)}(p)
        \sum_l\beta_l(k,\omega,p)\Bigl[z_{kp}Q_l(z_{kp})-\delta_{l,0}
        \Bigr]\ .
\label{legeself}
\end{equation}
The analytic properties of the self energy as well as the integral
equation are buried now into the $Q_l(z)$ function. It has a logarithmic
singularity at $|z|=1$,  it is a real function when $|z|>1$ and
complex when $|z|<1$. The singularity is integrable and thus the
q-space integration in Eq. (\ref{legecom}) and the p-space integration
in Eq. (\ref{legeself}) can be performed.

\newpage
\bibliography{papers}
\bibliographystyle{prsty}
\newpage
\begin{table}
\caption{Pressure dependence of the hydrodynamic effective mass from
various calculations and experiments.  The second column contains the
result of our microscopic calculation described in sections
\ref{single} and \ref{hydrmass}; the next column contains the
hydrodynamic effective mass as obtained from the fit (\ref{massfit})
proposed by theory to the experiments of
Ref. \protect\onlinecite{Yorozu93}.  Cols. 4 and 5 contain the
hydrodynamic effective mass as obtained from the linear extrapolation
and from the fit (\ref{massfit}) to the experiments of
Ref. \protect\onlinecite{SimonsLT21} and Col. 6 the results from the
fit to the magnetic susceptibility data.}
\vspace{0.2truein}
\begin{tabular}{cccccc}
$P$~~(atm)&\multicolumn{5}{c}{$m_H^*/m$}\\
& This work
& Ref. \protect\onlinecite{Yorozu93} (extrap.)
& Ref. \protect\onlinecite{SimonsLT21}
& Ref. \protect\onlinecite{SimonsLT21} (extrap.)
& magn. susc.\\
\hline
0   & 2.09 & 2.18 & 2.23$\pm$0.02 & 2.15 & 2.27 \\
5   & 2.22 & 2.31 & \\
10  & 2.34 & 2.44 & 2.52$\pm$0.02 & 2.39 & 2.42 \\
15  & 2.45 & 2.54 & \\
20  & 2.55 & 2.64 & 2.70$\pm$0.03 & 2.62 & 2.58 \\
\end{tabular}
\label{tab:hydmass}
\end{table}
\newpage
%
%
\begin{table}
\caption{The half width of the peak in the dynamic structure function
of the impurity excitation in the roton region for 1.8\AA$^{-1}<k
<2.4$\AA$^{-1}$.  The third column gives the distance 
the impurity excitation can propagate within its life time.}
\vspace{0.2truein}
\begin{tabular}{cccc}
$k$~~\AA$^{-1}$  &half width (K)& distance  (\AA) & \\
\hline
1.8 & 0.28 & 35.7& \\
1.9 & 0.63 & 16.2& \\
2.0 & 1.01 & 10.3& \\
2.1 & 1.55 & 6.84& \\
2.2 & 2.3 & 4.67& \\
2.3 & 3.1 & 3.51& \\
2.4 & 4.0 & 2.74& \\
\end{tabular}
\label{tab:penetra}
\end{table}
\newpage
%
\begin{table}
\caption{Pressure dependence of
the coefficients of the expansion (\ref{massfit}) for the
concentration dependence of the effective mass.
The expansion coefficients $a$, $b$, $c$, and $d$ are
from Ref.  \protect\onlinecite{mixprl}.}
\vspace{0.2truein}
\begin{tabular}{ccccc}
$P$~~(atm)  &  $a$ & $b$ & $c$ & $d$ \\
\hline
0  & 1.49 & 1.39 & -18.2 & 36.7 \\
5  & 1.07 & 3.00 & -22.6 & 40.2 \\
10 & 0.789 & 4.48 & -28.2 & 50.4 \\
15 & 0.501 & 6.17 & -36.1 & 66.8 \\
20 & 0.310 & 7.41 & -42.1 & 80.1 \\
\end{tabular}
\label{tab:massfit}
\end{table}
%
\newpage
\begin{table}
\caption{Pressure dependence of the parameters of the fit
(\ref{f0afit}) of the un-normalized Fermi liquid
parameter $(m/m^*)F_0^{\rm a}$ as obtained from the dynamic
calculation. \label{tab:F0afit}}
\vspace{0.2truein}
\begin{tabular}{ccccc}
$P~~$(atm)  & $ a $ & $ b $  & $ c $ & $  d$\\
\hline
   0  &  0.447 & -4.371  & 14.67 & -22.35 \\
   5  &  0.394 & -3.710  & 10.82 & -14.73 \\
  10  &  0.362 & -3.319  & 8.410 & -9.660 \\
  15  &  0.344 & -3.012  & 6.336 & -4.927 \\
  20  &  0.326 & -2.733  & 4.383 & -0.349 \\
\end{tabular}
\end{table}
\newpage
%
%
%
\newpage
\newpage
%
%
\newpage
%
%
\newpage
%
%
\newpage
%
%
\newpage
%
%
\newpage
%
%
%
%
\newpage
\newpage
%
%
\newpage
\newpage
%
%
\newpage
%
%
\newpage
%
%
\newpage
%
%

\end{document}